\documentclass[12pt]{article}
\usepackage{amsfonts,amssymb,amsmath,amscd}
\usepackage{amsmath,amssymb,epsfig}
\usepackage{graphicx}

\newcommand{\eps}{\epsilon}

\def\be{\begin{equation}}
\def\ee{\end{equation}}
\def\bear{\begin{eqnarray}}
\def\eear{\end{eqnarray}}

\def\half{{ \frac{1}{2} }}

\def\Id{{\mathbb{I}}}

\def\eps{{\epsilon}}
\title{Spectral networks and higher web-like structures}
\author{Natalia Saulina\\ Department of Physics and Astronomy,\\ Rutgers, The State University of New Jersey}
\begin{document}
\begin{titlepage}

\maketitle

\begin{abstract}

We derive traffic rule for spectral networks for $A_2$ theory for Riemann surface $C$ with punctures
and use it to study in details the moduli space $\mathcal{M}$ of flat $GL(3,\mathbb{C})$ connections on $\mathbb{P}^1$ with 3 full punctures.
We apply the simplified traffic rule to find the global description of $\mathcal{M}$
as the fibration with symplectic fibers
and to study the space of line defects in 4d theory corresponding to
the conformal point in the base of $\mathcal{M}.$
We define higher web-like structures, not previously discussed in the literature.
They give rise to line defects independent of the defects corresponding
to the traces of holonomies.
\end{abstract}
\end{titlepage}

\section{Introduction}
Moduli spaces of flat $GL(K,\mathbb{C})$ connections on Riemann surfaces
with punctures play a key role \cite{GMN-WKB, NRS} in the study of line defects in $\mathcal{N}=2, d=4$ supersymmetric theories. In particular, the algebra of functions on these
moduli spaces is relevant for computing the Operator Product Expansion of line
defects.

An explicit description of moduli spaces of flat $GL(K,\mathbb{C})$ connections 
and their algebras of functions was proposed in \cite{GMN-specnet,GMN-snakes} by means of spectral networks. Alternatively, Fock-Goncharov construction \cite{FG}
of moduli spaces of flat $PGL(K,\mathbb{C})$ connections was fruitfully used
\cite{Xie,Xie2}
in the literature on line defects. Traffic rule is a crucial tool in the description
of these moduli spaces. For $A_1$ theory the traffic rule
was formulated in \cite{Thurston,Penner} and used in \cite{GMN-WKB,GMN-snakes}. 
See also a review \cite{FG-review} on this subject.
%Note however, that depending on  a Riemann surface,
%the moduli space of flat $PGL(K,\mathbb{C})$ connections may only be a subspace
%of the corresponding moduli space of flat $GL(K,\mathbb{C})$ connections.
In this note we derive traffic rule for $A_2$ theory for general Riemann surface $C$(without boundary but with punctures) 
and use it to discuss in details the moduli space $\mathcal{M}$ of flat $GL(3,\mathbb{C})$ connections on $\mathbb{P}^1$ with 3 full punctures.

In the spectral network approach, one triangulates a Riemann surface $C$
 such that punctures serve as vertices of the triangles and considers branched cover $\Sigma$ with $\half K(K-1)$ branch points in each triangle. 
To explicitly write down a monodromy of a flat $GL(K,\mathbb{C})$ connection on $C$
around an arbitrary contour one needs to formulate the traffic rule.
  For $A_K$ theory with $K>1$
the prescription of how to cross branch cuts and edges of the triangles was missing, and we provide it  in Section 3 for $A_2$ theory. Then we pass to a simplified version
of spectral network which is sufficient to describe monodromies of homotopy classes of curves. Traffic rule for simplified spectral networks generalizes easily to
$A_K$ theory with $K>2$ (see Section 4).

We apply the simplified traffic rule to find the global description of $\mathcal{M}$
as the fibration with symplectic fibers given in (11)
and to study the space of line defects in 4d theory corresponding to
the conformal point in the base of $\mathcal{M}.$
We show that basic webs give rise to line defects that are linear combinations of defects arising from traces of holonomies. 
We define higher web-like structures, not previously discussed in the literature, and show that they give rise to line defects independent of the defects corresponding
to the traces of holonomies.

%for $A_2$ theory on $C=\mathbb{P}^1/\{v_1,v_2,v_3\}.$

This note is organized as follows. In Section 2 we review some basic facts about the moduli space $\mathcal{M}$ of flat $GL(3,\mathbb{C})$ connections on 
$C=\mathbb{P}^1/\{v_1,v_2,v_3\}.$ In Section 3 we provide traffic rule for $A_2$ theory for general $C$ (without boundary but with punctures) and use the traffic rule
to describe $\mathcal{M}$ by means of spectral networks and  to find the global description of $\mathcal{M}.$
In Section 4 we formulate the simplified traffic rule for $A_2$ theory  and generalize it for $A_K$ theory
with $K>2.$
In Section 5 we discuss the space of line defects at the conformal point. In Section 6 we give an explicit relation between the spectral network and Fock-Goncharov descriptions.

 \section{Flat $GL(3,\mathbb{C})$ connections on $\mathbb{P}^1$ with three punctures}
\label{sec:rev}
Moduli space $\mathcal{M}$ of flat $GL(3,\mathbb{C})$ connections on $C=\mathbb{P}^1/\{v_1,v_2,v_3\}$
can be parametrized by three $GL(3,\mathbb{C})$ matrices $M_1,M_2,M_3$
up to simultaneous conjugation $M_i\mapsto h M_i h^{-1}$ and a constraint
$$M_1M_2M_3=\Id.$$ These matrices are monodromies around the punctures
with some fixed based point. Counting independent parameters gives ten-dimensional
moduli space:
$$3\times 9- 9-8=10$$
where  we took into account that conjugation with $h=\lambda \Id$ lives the constraint invariant.

\begin{figure}
%\centering
\includegraphics[width=1.2\textwidth]{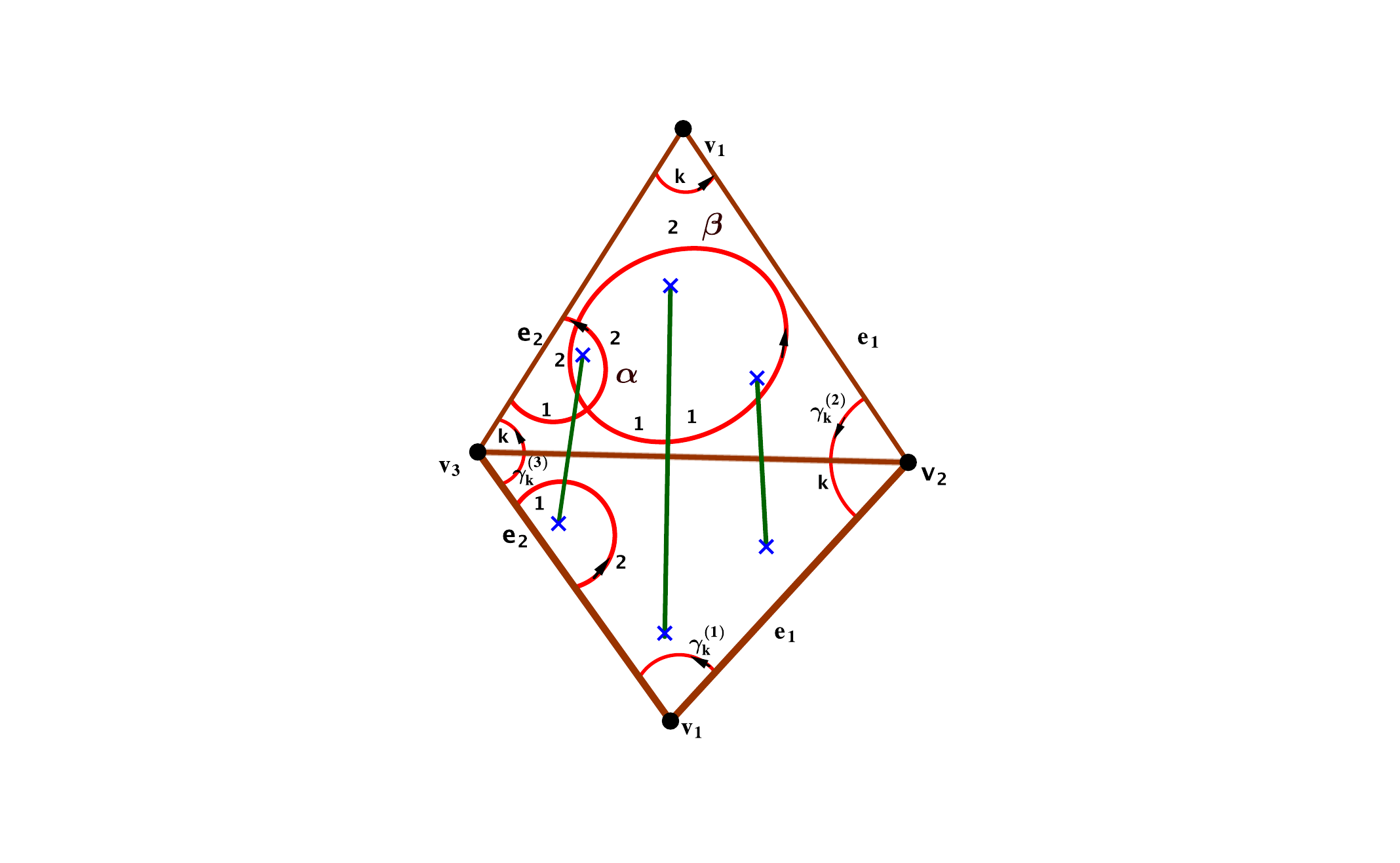}
\vspace{-2cm}
\caption{Basis of non-trivial 1-cycles on $\Sigma$. Numbers indicate sheets of the cover.}
\label{fig:cycles}
\end{figure}

$\mathcal{M}$ is a Poisson manifold and the Darboux coordinates $X_{\gamma_i}$can be identified, following the general proposal
in \cite{GMN-specnet}, with holonomies of  a flat abelian connection around basic non-trivial
1-cycles $\gamma_i$ on $3:1$ branched cover $\Sigma \mapsto C.$
There are three branch points of the type 12, 23,12 respectively in each triangle
of a chosen triangulation of $C.$
 The Poisson structure is simply
$$\{X_{\gamma_i},X_{\gamma_j}\}=\langle \gamma_i,\gamma_j\rangle X_{\gamma_i}\,X_{\gamma_j}$$
where $\langle \cdot, \cdot \rangle$ is the intersection pairing between the 1-cycles.
Applying Hurwitz formula, we find that for $C=\mathbb{P}^1/\{v_1,v_2,v_3\}$ the cover
$\Sigma$
is $\mathbb{T}^2$ with nine punctures. The basis of 1-cycles on $\Sigma$ is shown
in Figure \ref{fig:cycles}. 

The only non-zero Poisson bracket is
$$\{X_{\alpha}, X_{\beta}\}=X_{\alpha}\, X_{\beta}$$
The 1-cycles $\gamma^{(a)}_k$ go around nine punctures $p^{(a)}_k$  on $\Sigma$ 
which are lifts to the $k=1,2,3$ sheet of the three punctures $v_a$ on $C$. The corresponding Darboux coordinates
$X_{\gamma^{(a)}_k}$ are in the center of the Poisson algebra and satisfy
$$\prod_{a=1}^3\prod_{k=1}^3 X_{\gamma^{(a)}_k}=1.$$ $\mathcal{M}$ has the structure of a fibration  over the manifold parametrized by $X_{\gamma^{(a)}_k}$ with  a fiber being a complex symplectic manifold with coordinates $X_{\alpha},X_{\beta}.$
By fixing conjugacy classes of monodromies around punctures, one fixes a point
in the base and works with this symplectic manifold.

\section{$\mathcal{M}$ from spectral network}
\label{sec:specnet}
Here we describe $\mathcal{M}$ by means  of spectral networks. In Section 3.1 we discuss the details of the traffic rule required to construct
a flat connection on a trivialized $GL(3,\mathbb{C})$ bundle on a Riemann surface $C$ with punctures.
In Section 3.2 we apply this to compute monodromies around
punctures for $C=\mathbb{P}^1/\{v_1,v_2,v_3\}$ and to provide the global description of
$\mathcal{M}.$
\subsection{Traffic rule for $A_2$ theory}
\label{sec:traffic}
The notion of a spectral network was introduced in \cite{GMN-specnet} as a tool
for the computation of BPS degeneracies in $\mathcal{N}=2$ theories. In the same paper, it was also
proposed to use spectral networks to construct flat  connections on a trivialized $GL(K,\mathbb{C})$ bundle
on a Riemann surface $C$ with punctures. In this construction, one triangulates $C$
 such that punctures serve as vertices of the triangles and considers branched cover $\Sigma$ with $\half K(K-1)$ branch points in each triangle. $S$-walls of various type
 start at branch points and go to the punctures in such a way that in each triangle
 there is a level K lift of $AD_1$ theory. In \cite{GMN-snakes} two types of level K lifts were introduced - Yin and Yang. In Section \ref{sec:specnet} we work with the Yin type, but in Section \ref{sec:simple} we describe simplified traffic rule for both types.
Given a triangulation of $C$ and a trivialization of the $GL(K,\mathbb{C})$ bundle,  an assignment of Yin and Yang networks to the triangles gives a chart
of an atlas with which we can describe $\mathcal{M}.$ 

 $GL(K,\mathbb{C})$ matrices that
describe crossing S-walls (from right to left) were found in \cite{GMN-snakes}. For example for $K=3,$ these matrices are 
$$S_{12}=\begin{pmatrix}1 & 1 & 0\cr 0 &1 &0\cr 0 & 0&1\end{pmatrix},\quad S_{23}(r)=\begin{pmatrix}r & 0 & 0\cr 0 & 1 & 1\cr 0 & 0 &1\end{pmatrix},
\quad S_{21}(r)=\begin{pmatrix}1& 0& 0\cr-1 &1&0\cr0&0&r\end{pmatrix},\quad
S_{32}=\begin{pmatrix}1 & 0 & 0\cr 0 &1 &0\cr 0 & -1&1\end{pmatrix}.$$
 Here $r$ is a complex parameter for each triangle.
 
 To explicitly write down a monodromy of a flat  connection on a trivialized $GL(K,\mathbb{C})$ bundle on $C$
 around an arbitrary contour $\mathcal{P}$ one needs to formulate the traffic rule
 i.e. prescribe what $GL(K,\mathbb{C})$ transformation corresponds to crossing of
 the path $\mathcal{P}$ with $S$-walls as well as with various branch cuts
and edges of the triangles. For $A_1$ theory the complete traffic rule
was formulated in \cite{GMN-WKB,GMN-snakes}.  For $A_K$ theory with $K>1$
the prescription of how to cross branch cuts and edges was missing, and we provide it
 in this note.

\begin{figure}
%\centering
\includegraphics[width=1.4\textwidth]{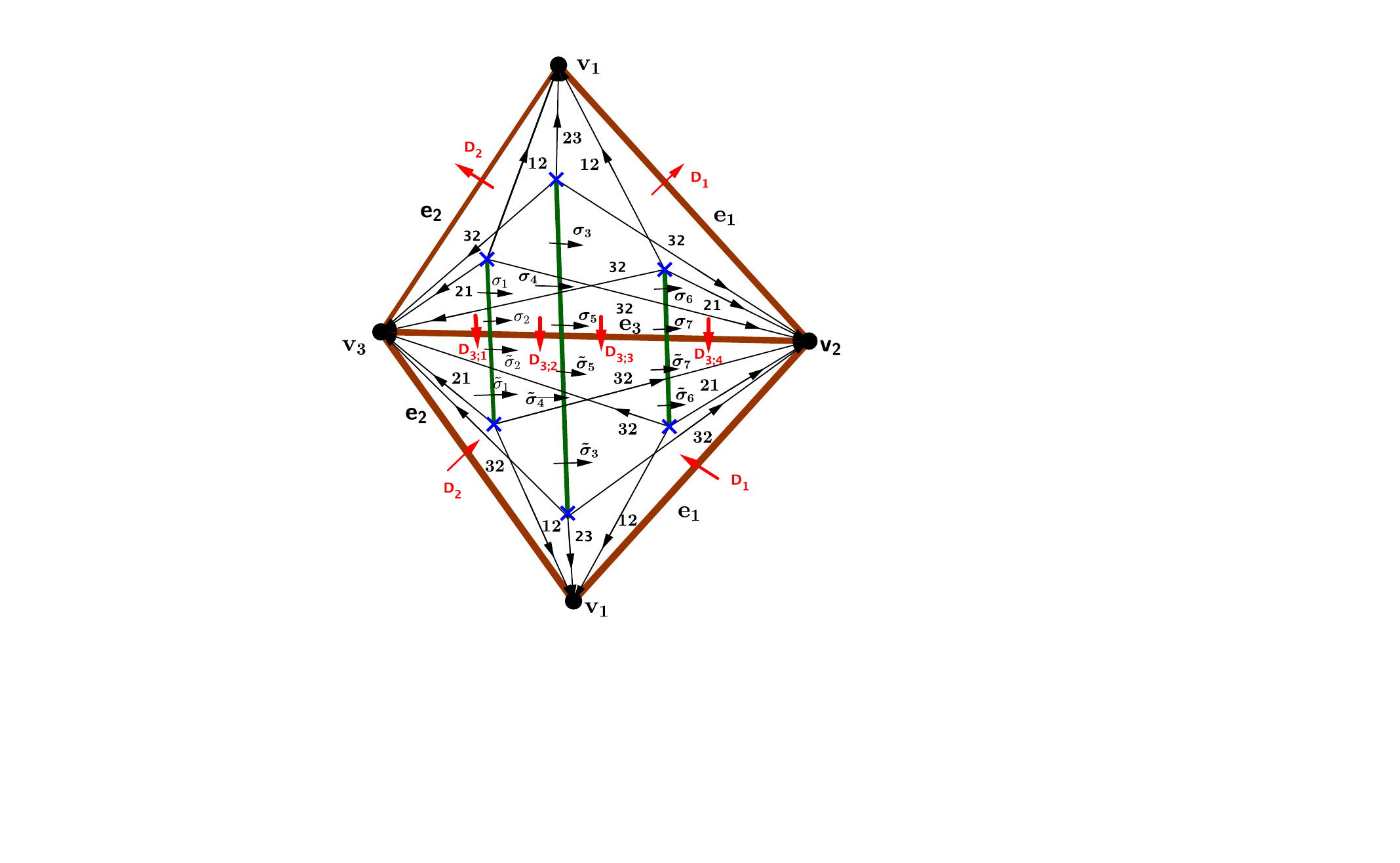}
\vspace{-5cm}
\caption{Traffic rule for $A_2$ theory on $\mathbb{P}^1/\{v_1,v_2,v_3\}$. The upper (lower) triangle is characterized by complex parameter $r_2$($r_1$).}
\label{fig:traffic}
\end{figure}

\begin{figure}
\begin{center}
\includegraphics[width=1.6\textwidth]{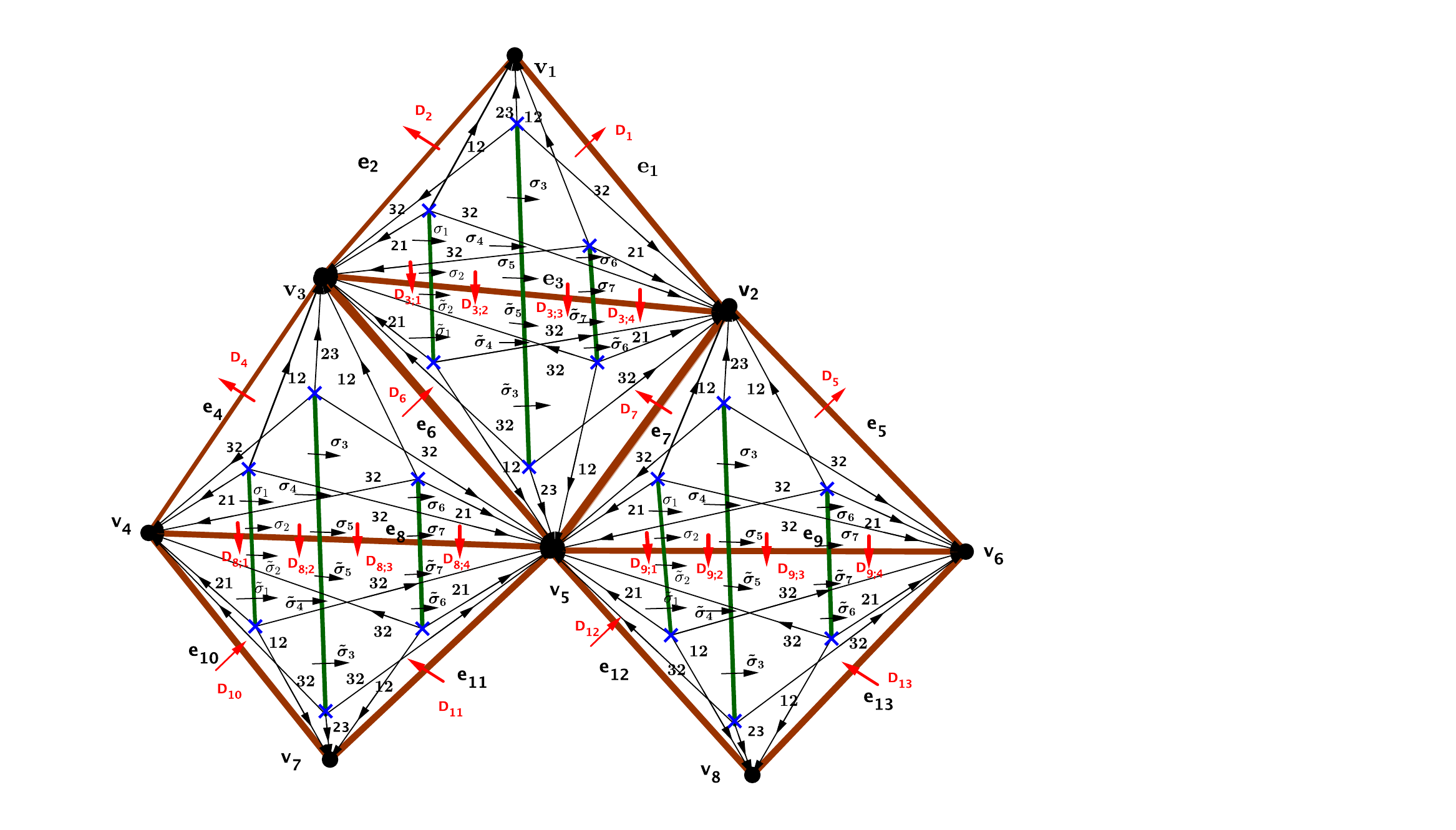}
\caption{Traffic rule for  minimal spectral network in general $A_2$ theory. It is implicitly
assumed that matrices of crossing branch cuts depend on the parameter $r$
for the appropriate triangle. }
\label{fig:traffic_gen}
\end{center}
\end{figure}

It is nontrivial to observe that the traffic rule depends on a choice of branch cuts. Moreover, each branch cut is split into segments by $S$-walls intersecting it.
To ensure the flatness of the connection, implying that monodromy is an identity matrix around any contractible
cycle on $C,$ we have to assign certain matrices $\sigma_I$ to crossing various segments $I$ of the branch cuts. It is sufficient to determine $\sigma_I$ for
$C=\mathbb{P}^1/\{v_1,v_2,v_3\}$ (as in Figure \ref{fig:traffic}) since similar matrices, though depending on the parameter $r$ for the appropriate triangle,  describe crossing the branch cuts in the other triangles for more general
$C$ (see Figure \ref{fig:traffic_gen}). 

For technical reasons we choose\footnote{With this choice, monodromies around punctures have simpler analytic expressions for their eigenvalues.} branch cuts so that each edge is either not intersected
by branch cuts or is intersected precisely by $\half K(K-1)$ branch cuts.  
If the edge of the triangle is not intersected by branch cuts, we assign
diagonal matrix to the crossing of this edge. If the edge $j$ is crossed by the cuts, we assign diagonal crossing matrix $D_{(j;1)}$ to the left-most segment of the edge, while crossing  matrices $D_{(j;a)}$ for all 
other segments of this edge (i.e. $a=2,\ldots, \half K(K-1)+1$ ) are determined in terms of   $D_{(j;1)}$ and matrices $\sigma_I$
of crossing the branch cuts. It turns out  that  $D_{(j;a)} \, a=2,\ldots, \half K(K-1)$  are non-diagonal matrices.

%\subsection{Sphere with three punctures}
In Figure \ref{fig:traffic} we give an example of the traffic rule for $A_2$ theory on 
$C=\mathbb{P}^1/\{v_1,v_2,v_3\}.$
To ensure the flatness of the connection, implying that monodromy is identity matrix around any contractible
cycle on $\mathbb{P}^1/\{v_1,v_2,v_3\},$ we have to assign the following matrices to crossing various segments 
of the branch cuts 

$$\sigma_2=S^{-1}_{12}S^{-1}_{21}(r_2)S^{-1}_{32},\quad \tilde \sigma_2=
S_{12}S_{21}(r_1)S_{32}$$
\be \label{sigma-mat}\sigma_5=S^{-1}_{32}S^{-1}_{23}(r_2)S^{-1}_{12}S^{-1}_{32}S_{12},
\quad \tilde \sigma_5=
S_{32}S_{23}(r_1)S_{12}S_{32}S^{-1}_{12}\ee
$$\sigma_7=S^{-1}_{32}S^{-1}_{21}(r_2)S^{-1}_{12},\quad \tilde \sigma_7=
S_{32}S_{21}(r_1)S_{12}$$
For example, we compute
$$\sigma_7\sigma_5\sigma_2=M_a^{-1}(r_2)M_b^{-1}(r_2)S^{-1}_{32}S^{-1}_{21}(r_2)S^{-1}_{32}=M_a^{-1}(r_2)M_b^{-1}(r_2)M_a^{-1}(r_2).$$
Here we used the matrices of crossing cables from \cite{GMN-snakes}
$$M_a(r)=M_c(r)=S_{32}S_{21}(r)S_{32},\quad M_b(r)=S_{12}S_{23}(r)S_{12}.$$
The other crossing branch-cuts matrices are determined as follows:
\be
\label{other-sigma}
\sigma_1=S^{-1}_{32}\sigma_2S_{32}\quad
\sigma_4=S^{-1}_{32}\sigma_5 S_{32}\quad \sigma_3=\sigma_5\quad
\sigma_6=S_{32}\sigma_7S^{-1}_{32}\ee
$$\tilde\sigma_1=S_{32}\tilde\sigma_2S^{-1}_{32}\quad
\tilde\sigma_4=S_{32}\tilde\sigma_5S^{-1}_{32}\quad \tilde\sigma_3=\tilde\sigma_5\quad
\tilde\sigma_6=S^{-1}_{32}\tilde\sigma_7S_{32}$$
We further introduce diagonal matrices $D_{(1)}$ and $D_{(2)}$ for crossing edges $e_1$ and $e_2$ which are not cut into segments by branch cuts. For the edge $e_3,$ we introduce diagonal matrix $D_{(3;1)}$ for the crossing of the left-most segment of $e_3$
and compute from the requirement of flatness of the $GL(3,\mathbb{C})$ connection:

\be
\label{d3}
D_{(3;2)}=\tilde\sigma_2 D_{(3;1)}\sigma_2^{-1},\quad
D_{(3;3)}=\tilde\sigma_5 D_{(3;2)}\sigma_5^{-1},\quad
D_{(3;4)}=\tilde\sigma_7 D_{(3;3)}\sigma_7^{-1}\ee

Below we give explicit matrices used in crossing segment of branch cuts and edges:
$$\sigma_1=\begin{pmatrix}0 & -1 &0\cr 1&1&0\cr1& 1&r_2^{-1}\end{pmatrix},\quad
\sigma_2=\begin{pmatrix}0 & -1 &0\cr 1&1&0\cr 0 & r_2^{-1}&r_2^{-1}\end{pmatrix}, \quad
\sigma_3=\sigma_5=\begin{pmatrix}r_2^{-1} & 0 &0\cr 0&0&-1\cr 0 & 1& 0\end{pmatrix}
$$
$$\sigma_4=\begin{pmatrix}r_2^{-1} & 0 &0\cr 0&1&-1\cr 0 & 2& -1\end{pmatrix},\quad
\sigma_6=\begin{pmatrix}1 & -1 &0\cr 1&0&0\cr 0 & r_2^{-1}&r_2^{-1}\end{pmatrix}, \quad \sigma_7=\begin{pmatrix}1 & -1 &0\cr 1&0&0\cr 1 & 0 &r_2^{-1}\end{pmatrix}$$

$$\tilde \sigma_1=\begin{pmatrix}0 & 1 &0\cr -1&1&0\cr1& -1&r_1\end{pmatrix},\quad
\tilde\sigma_2=\begin{pmatrix}0 & 1 &0\cr -1&1&0\cr 0 & -r_1&r_1\end{pmatrix}, \quad
\tilde \sigma_3=\tilde \sigma_5=\begin{pmatrix}r_1 & 0 &0\cr 0&0&1\cr 0 & -1& 0\end{pmatrix}
$$
$$\tilde \sigma_4=\begin{pmatrix}r_1 & 0 &0\cr 0&1&1\cr 0 & -2& -1\end{pmatrix},\quad
\tilde\sigma_6=\begin{pmatrix}1 & 1 &0\cr -1&0&0\cr 0 & -r_1&r_1\end{pmatrix}, \quad \tilde \sigma_7=\begin{pmatrix}1 & 1 &0\cr -1&0&0\cr 1 & 0 &r_1\end{pmatrix}$$

$$D_{(i)}=\begin{pmatrix}x_i & 0 &0\cr 0&y_i&0\cr 0 & 0 &z_i\end{pmatrix},\quad i=1,2$$
$$D_{(3;1)}=\begin{pmatrix}x_3 & 0 &0\cr 0&y_3&0\cr 0 & 0 &z_3\end{pmatrix},\quad
D_{(3;2)}=\begin{pmatrix}-y_3 & 0 &0\cr -
(x_3 + y_3)&-x_3&0\cr (y_3 + z_3)r_1 & 0 &z_3r_1r_2\end{pmatrix}$$

$$D_{(3;3)}=\begin{pmatrix}-y_3 r_1 r_2& 0& 0\cr (y_3   +  z_3) r_1 r_2& -z_3 r_1 r_2& 0\cr (x_3  +  y_3) r_2 &
   0& x_3\end{pmatrix},\quad D_{(3;4)}=r_1r_2\begin{pmatrix}z_3 & 
0& 0\cr 0 & 
  y_3  & 0\cr 0 & 0 &x_3 \end{pmatrix}$$

The traffic rule for $A_2$ theory on a more general $C$ is depicted on Figure
 \ref{fig:traffic_gen}, where
 we implicitly
assume that matrices of crossing branch cuts depend on the parameter $r$
for the appropriate triangle. Figure \ref{fig:traffic_gen} shows three rhombi, each built of a pair of triangles connected by three branch cuts. Finding traffic rule for a general $C,$
i.e. for any number of such rhombi,
is very similar. This is the case due to our choice of branch cuts - three branch cuts are
contained within each rhombus and no branch cuts go between different rhombi.
So that one uses closed contours  within each rhombus to determine
matrices of crossing the branch cuts contained in the rhombus. Meanwhile, crossing
the edges between different rhombi is described simply by diagonal matrices. 

We introduce diagonal matrices $D_{(1)},D_{(2)},
D_{(4)},D_{(5)},D_{(6)},D_{(7)},D_{(10)},
D_{(11)},D_{(12)},D_{(13)}$ for crossing the edges not split into segments by branch cuts as well as diagonal matrices  $D_{(3;1)},D_{(8;1)},D_{(9;1)}$ for the left-most
segments of the edges $e_3,e_8,e_9$
and compute similar to (\ref{d3})
$$D_{(8;2)}=\tilde\sigma_2(r_3) D_{(8;1)}\sigma_2(r_4)^{-1},\quad
D_{(8;3)}=\tilde\sigma_5(r_3) D_{(8;2)}\sigma_5(r_4)^{-1},\quad
D_{(8;4)}=\tilde\sigma_7(r_3) D_{(8;3)}\sigma_7(r_4)^{-1}$$
$$D_{(9;2)}=\tilde\sigma_2(r_5) D_{(9;1)}\sigma_2(r_6)^{-1},\quad
D_{(9;3)}=\tilde\sigma_5(r_5) D_{(9;2)}\sigma_5(r_6)^{-1},\quad
D_{(9;4)}=\tilde\sigma_7(r_5) D_{(9;3)}\sigma_7(r_6)^{-1}$$
where we indicate the dependence on the appropriate $r_i$ parameters.

\subsection{Monodromies around punctures}
\begin{figure}
\centering
\includegraphics[width=1.4\textwidth]{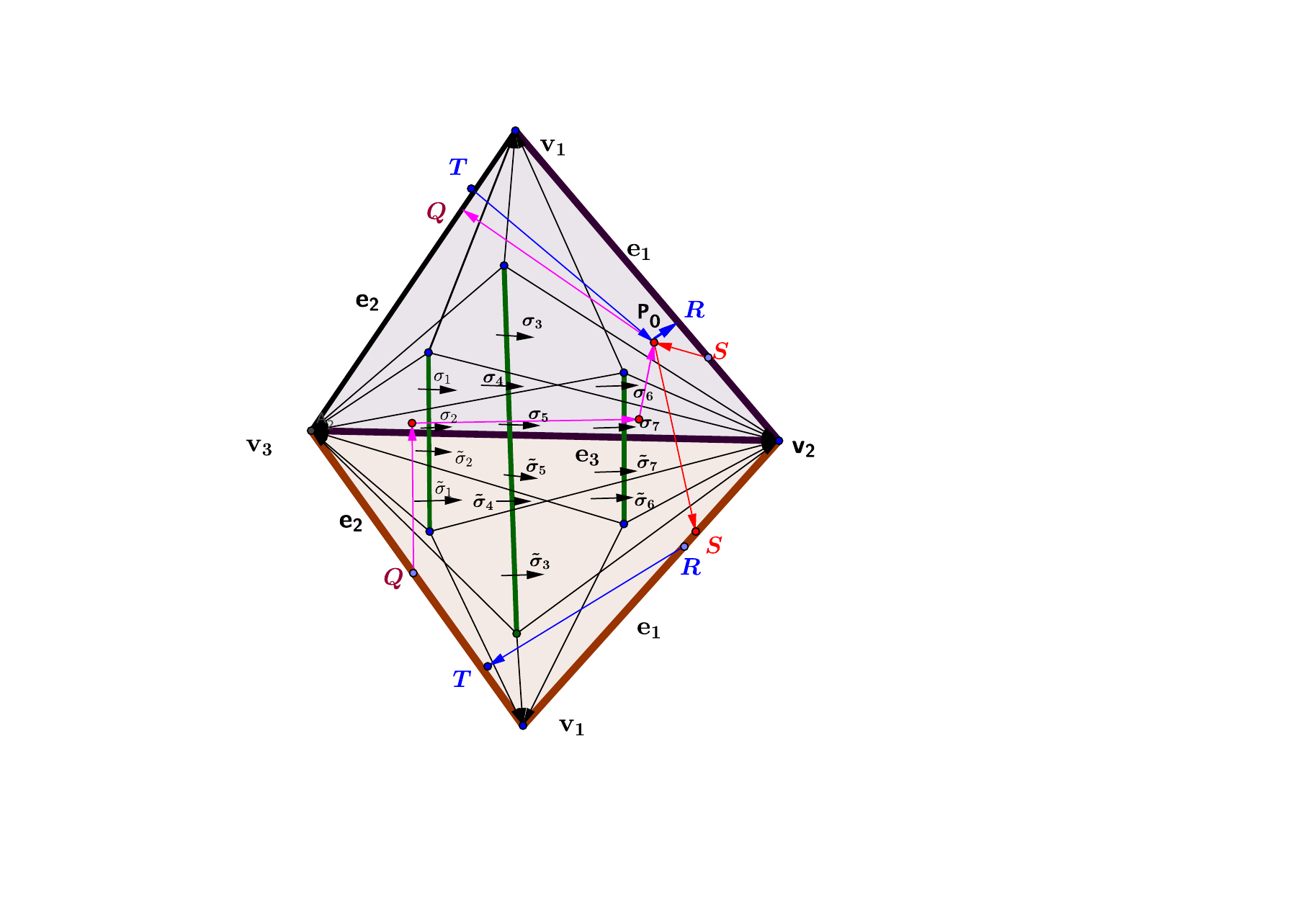}
\vspace{-2cm}
\caption{Monodromies around punctures with base point $P_0$}
\label{fig:monodromy}
\end{figure}

We are interested in flat $GL(3,\mathbb{C})$ connections with fixed conjugacy
classes of the monodromies around the punctures. Let $P_0$ be a base point
as shown in Figure \ref{fig:monodromy}. This choice will allow to introduce simplified
network in Section \ref{sec:simple}.
The closed paths, starting and ending at $P_0,$ around $v_1,v_2,v_3$ are shown in Figure \ref{fig:monodromy}.

We use traffic rule formulated in the previous section to compute the corresponding monodromies

\be \label{mon_pun} M_{P_0}(v_1)=M^{-1}_b(r_2)D^{-1}_{(2)}M^{-1}_b(r_1)D_{(1)},\quad
M_{P_0}(v_2)=D^{-1}_{(1)}M^{-1}_a(r_1)D_{(3;4)}M^{-1}_a(r_2)\ee
$$M_{P_0}(v_3)=M_a(r_2) W(r_2) D^{-1}_{(3;1)} M^{-1}_a(r_1)
D_{(2)}M_b(r_2)$$
where we denote $W(r_2)=\sigma_7 \sigma_5 \sigma_2.$
These monodromies satisfy
\be \label{constr}M_{P_0}(v_3)\,M_{P_0}(v_1)\,M_{P_0}(v_2)=\Id\ee

Eigenvalues of these monodromies are 
$$Eigen\Bigl(M_{P_0}(v_1)\Bigr)=\Bigl\{{z_1\over z_2}, {y_1\over y_2},{x_1\over x_2 \, r_1\, r_2}\Bigr\}$$
$$Eigen\Bigl(M_{P_0}(v_2)\Bigr)=\Bigl \{{x_3\over z_1}, {r_1\,r_2\, y_3\over y_1},{r_1\,r_2\, z_3\over x_1}\Bigr \}$$
$$Eigen\Bigl(M_{P_0}(v_3)\Bigr)=\Bigl \{{z_2\over z_3\, r_1\,r_2}, {y_2\over y_3},{x_2\over x_3}\Bigr \}$$

Let us fix these eigenvalues in terms of masses  $k_j\, j=1,\ldots,9$
such that $\prod_{j=1}^9k_j=1.$ Then, we find
$$r_1r_2=k_2k_5k_8,\, z_1=k_1z_2,\, z_2=x_2 (k_2k_3k_5k_6k_7k_8),\,
y_1=k_2y_2,\, x_1=x_2 (k_2k_3k_5k_8)$$
$$x_3=k_4z_1,\, y_3={y_1\over k_2k_8},\, z_3={x_1k_6\over k_2k_5k_8} $$
It follows that, for fixed masses $k_j,$ monodromy around any closed path is a function of $r_2$ and $s={y_2\over x_2}.$

For example,
$$Tr\Bigl(M^2(v_1)\,M(v_2)\Bigr)=(k_1+k_2+k_3)
\left({1\over k_7}+{1\over k_8}+{1\over k_9}\right)+r_2\left(k_1k_3k_4+{k_3\over k_8}
+{1\over k_5k_8k_9}\right)+$$
$${1\over r_2}\left(k_1k_2k_5+{k_2\over k_7}
+{1\over k_4k_7k_9}\right)+
s\left(k_1k_2k_4\left(1+{k_9\over k_8}\right)+{(k_1+k_2)\over k_8}
+{1\over k_5k_7k_8}+{1\over k_6k_7k_8}\right)+$$
$${1\over s}\left({k_3 \over k_9}+{1\over k_4 k_7 k_9}\right)+{r_2\over s}\left({k_3\over k_9}\right)+{s\over r_2}\left(k_1k_2k_5+{k_2\over k_3k_6k_7k_8}
+{k_2\over k_7}\right)+$$
$$r_2 s\left({k_1k_4\over k_5k_8}+{1\over k_5k_8^2}+{k_1k_3k_4k_9\over k_8}\right)+
{1\over r_2 s}\left({1\over k_4 k_7 k_9}\right)+{s^2\over r_2}\left({k_2\over k_3k_6k_7k_8}\right)+s^2r_2\left({k_1k_4k_9\over k_5k_8^2}\right)+s^2\left({2k_1k_2k_4k_9\over k_8}\right)$$
Let us clarify the relation between coordinates $r_2, s, k_j$ on $\mathcal{M}$ and
Darboux coordinates reviewed in Section \ref{sec:rev}. The masses are identified
with the center of the Poisson algebra:
$$X_{\gamma^{(1)}_j}=k_j,\quad X_{\gamma^{(2)}_j}=k_{3+j},\quad X_{\gamma^{(3)}_j}=k_{6+j}\quad j=1,2,3.$$
It was shown in \cite{GMN-snakes} that $X_{\beta}=r_2.$ Finally, $s={y_2\over x_2}$ is naturally identified
with $X_{\alpha}$ since 1-cycle $\alpha$ intersects\footnote{Recall that $D^{(2)}=diag(x_2, y_2, z_2)$ describes crossing the edge $e_2$.}  the edge $e_2$ once on sheet 1 and 
once on sheet 2 (with opposite orientation). This identification with Darboux coordinates implies
\be \label{nonzero}\{ r_2,s\}=r_2\,s.\ee

 So far we worked with local coordinates, but it turns out that
$\mathcal{M}$ can be globally defined by 2 equations in 12 variables.
Let us define
\be \label{def-var}R_1=Tr\Bigl(M^2_{P_0}(v_1)M_{P_0}(v_2)\Bigr)-3,\quad
R_2=Tr\Bigl(M^2_{P_0}(v_2)M_{P_0}(v_1)\Bigr)-3\ee
\be \label{def-varii}L={1\over 3}Tr\Bigl(M^2_{P_0}(v_2)M_{P_0}(v_1)
M^{-1}_{P_0}(v_2)M^{-1}_{P_0}(v_1)\Bigr)-{1\over 3}Tr\Bigl(M^2_{P_0}(v_1)M_{P_0}(v_2)
M^{-1}_{P_0}(v_1)M^{-1}_{P_0}(v_2)\Bigr)\ee
\be \label{def-variii}f_a^j=Tr\Bigl(M^j_{P_0}(v_a)\Bigr) \quad a=1,2,3; \quad j=1,2,3.\ee
The first of the two equations defining $\mathcal{M}$
\be \label{def-eq} \Bigl(\frac{1}{3}f_1^3+\frac{1}{6}(f_1^1)^3-\half
f_1^1f_1^2\Bigr)\Bigl(\frac{1}{3}f_2^3+\frac{1}{6}(f_2^1)^3-\half
f_2^1f_2^2\Bigr)\Bigl (\frac{1}{3}f_3^3+\frac{1}{6}(f_3^1)^3-\half f_3^1f_3^2\Bigr)=1\ee
reflects the fact that $det\Bigl(M_{P_0}(v_1)\Bigr) det\Bigl(M_{P_0}(v_2)\Bigr) det\Bigl(M_{P_0}(v_3)\Bigr)=1.$
The second equation states
\be \label{eq-fiber}L^2-u_1LR_1-u_2LR_2-u_3L-g_1 R_1^2R_2^2-g_2R_1^3-g_3LR_1R_2-g_4R_1^2R_2-g_5R_2^3-g_6R_2^2R_1\ee
$$-g_7R_1^2-g_8R_2^2-g_9R_1R_2-g_{10}R_1-g_{11}R_2-g_{12}=0$$
where $u_1,\ldots,g_{12}$ are functions of $f_a^j.$ 
To get this equation we looked at the powers of $r_2$ and $s$ in each monomial
$L^{n_1}R_1^{n_2}R_2^{n_3}$ with non-negative integers $n_1,n_2,n_3.$
Namely, we used that $L$ contains powers of $s$ not higher than $s^3$  and not lower than $s^{-3},$ and powers of $r_2$ not higher than $r_2^2$ and not lower than $r_2^{-2}.$
We further used that $R_1(R_2)$ has powers of $s$ not higher than $s^2(s)$
and not lower than $s^{-1}(s^{-2}),$ and powers of $r_2$ not higher than $r_2$ and not lower than $r_2^{-1}.$So the procedure to get coefficients $g_i$  in (\ref{eq-fiber})
is to compute $L^2$ and first look at the monomial $s^6r_2^4.$ Besides $L^2$
only terms $g_1R_1^2R_2^2$ and $g_3LR_1R_2$ have such a monomial,
so this gives equation for $g_1,g_3.$ Then one looks at $s^6r_2^3$ etc.
In Appendix we give coefficients $g_i$ for a simple case when all $f_a^j$ are parametrized by a single variable $n.$

From (\ref{def-eq}) and (\ref{eq-fiber}) we see that $\mathcal{M}$ has the structure of
the fibration with the base defined by (\ref{def-eq}) and the fiber by (\ref{eq-fiber}).
Note that the fiber is a non-compact complex surface in $\mathbb{C}^3.$
We discuss in details the special point in the base, the conformal point, in Section 5.
At the conformal point the fiber is singular
$$L^2=R_1^2R_2^2-4(R_1+R_2)^3,$$
but away from this point the singularity is deformed, as we demonstrate in Appendix.

For the choice of the base point $P_0$ as in Figure \ref{fig:monodromy}, monodromies around punctures (\ref{mon_pun}) can be expressed using simplified spectral network
shown in Figure \ref{fig:simple}. Also, any closed curve with base point $P_0$ which can be drawn on Figure \ref{fig:monodromy}
is homotopic  to a closed curve that can be drawn on a simplified Figure \ref{fig:simple}.
We formulate the rules of the simplified network for general $C$ in Section 4.

\section{Simplified spectral network for general $C$}
\label{sec:simple}
 \begin{figure}
%\begin{center}
\includegraphics[width=\textwidth]{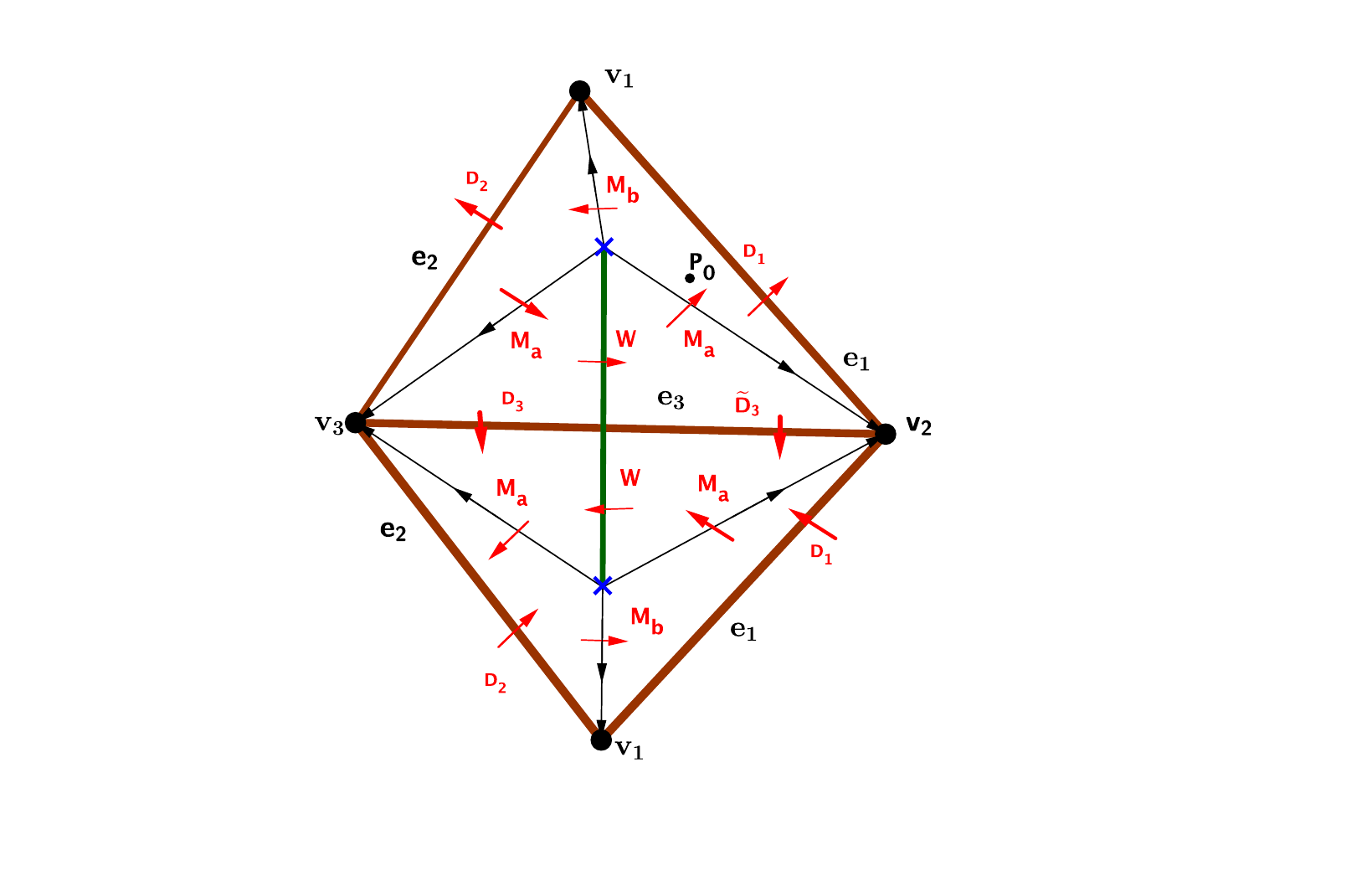}
\caption{Simplified spectral network for  $A_2$ theory on $\mathbb{P}^1/\{v_1,v_2,v_3\}$. $M_a(r), M_b(r)$ and $W(r)$ depend on the type of the spectral network -Yin or Yang- in a given triangle. }
\label{fig:simple}
\end{figure}
Any Riemann surface $C$ with punctures and without boundary can be triangulated by even number of triangles, and in each pair of triangles one can draw the simplified network shown in Figure \ref{fig:simple}. Moreover, in each triangle one may consider either Yin or Yang type of spectral
network. In Section \ref{sec:specnet} we discussed in details Yin type networks but the story is
similar for the Yang type.
In a simplified network in Figure \ref{fig:simple}, the following matrices are used for Yin-type  
 $$M_b(r)=S_{12}\, S_{23}(r)\,S_{12}=
 \begin{pmatrix}r & 1 + r& 1\cr 0& 1& 1\cr 0& 0& 1\end{pmatrix}, \quad
  M_a(r)=S_{32}\,S_{21}(r)\,S_{32}=\begin{pmatrix}1& 0& 0\cr -1& 1& 0\cr 1& -(1 + r)& r\end{pmatrix},$$
  while for  Yang-type  
  $$M_b(r)=S_{21}(r)\,S_{32}\,S_{21}(r)=
  \begin{pmatrix}1& 0& 0\cr -2& 1& 0\cr r& -r& r^2\end{pmatrix}, \quad
  M_a(r)=S_{23}(r)\,S_{12} \,S_{23}(r)=\begin{pmatrix}r^2& r& r\cr 0& 1& 2\cr 0& 0& 1\end{pmatrix}.$$
 In both types of triangles 
 \be \label{def_W}W(r)=M_a^{-1}(r)M_b^{-1}(r)M_a^{-1}(r).\ee
 While in the Yin-type triangle this gives an off-diagonal matrix
$$ W^{Yin}(r)={1\over r}\begin{pmatrix} 0 & 0 &1\cr 0& -1& 0\cr 1 & 0 & 0\end{pmatrix},$$
 this is not so in the Yang-type triangle
 $$W^{Yang}(r)=\left(
\begin{array}{ccc}
 \frac{2 r^2-2 r^3}{r^6} & \frac{-r^5+3 r^4-2 r^3}{r^6} & \frac{2 r^5-4 r^4+3 r^3}{r^6} \\
 \frac{2 r^4-2 r^3}{r^6} & \frac{r^6-4 r^5+2 r^4}{r^6} & \frac{-2 r^6+6 r^5-4 r^4}{r^6} \\
 \frac{1}{r^3} & \frac{r^5-r^4}{r^6} & \frac{2 r^4-2 r^5}{r^6} \\
\end{array}
\right).$$
% Figure \ref{fig:simple} shows simplified network for $C=\mathbb{P}^1/\{v_1,v_2,v_3\}$
 %but the story obviously generalizes, similar to how Figure \ref{fig:traffic_gen} generalizes Figure \ref{fig:traffic}. Namely, 
 The simplified spectral network can be generalized for $A_K$ theory with $K>2$ by taking
 cable matrices $M_a(r),M_b(r),M_c(r)$  given in \cite{GMN-snakes} and computing $W(r)$
 from (\ref{def_W}). 
 %For example for Yin-type network
 %$$M_a=\alpha_1\alpha_2\ldots\alpha_{K-2}\alpha_{K-1}$$
 %with
% $$\alpha_{K-1}=S_{K,K-1},\alpha_{K-2},\quad \alpha_{K-2}=S_{K,K-1}S_{K-1,K-2}, \ldots,
 %\alpha_2=S_{K,K-1}S_{K-1,K-2}\ldots S_{3,2},\quad  \alpha_2=S_{K,K-1}S_{K-1,K-2}\ldots S_{3,2} S_{2,1}$$
 
 \section{Space of line defects at the conformal point}
 Here we work at the conformal point in the base of $\mathcal{M}$ i.e. for
$f_a^j=3$ for $a=1,2,3$ and $j=1,2,3.$ We investigate if the algebra of  line
defects in 4d theory is all of Functions on the fiber at the conformal point.
For technical reasons, we do most of the computations in the Yin-Yin chart with coordinates $(r_2,s)\in \mathbb{C}^*\times \mathbb{C}^*$ (see however expressions of the generators $R_1,R_2,L$ in
the Yang-Yang chart at the end of this section). 
We conclude that traces of holonomies and webs {\it together}  do not generate
all of Functions of $(r_2,s)\in \mathbb{C}^*\times \mathbb{C}^*.$ 
%characterized by invariance
%(up to a possible sign) under $r_2\leftrightarrow r_2^{-1},s\leftrightarrow s^{-1}.$
It is interesting that higher web-like structures, not previously discussed in the literature,
correspond to line defects with vevs that are functionally independent of the vevs of
defects arising from traces of holonomies. 

Let us solve (\ref{constr}) as
$$M_{P_0}(v_3)=M^{-1}_{P_0}(v_2)M^{-1}_{P_0}(v_1),$$
simplify our notations $M_j=M_{P_0}(v_j)$
and consider traces of words built from `letters'  $M_1^{\pm 1},M_2^{\pm 1}.$ We found that the space of these traces is generated by
$R_1,R_2,L$ defined in (\ref{def-var}-\ref{def-varii}). At the conformal point in the base
they  satisfy the relation
\be \label{R_1R_2L}L^2=R_1^2R_2^2-4(R_1+R_2)^3.\ee
Let us denote line defects in 4d theory corresponding to certain linear combinations of traces of words
as follows
\be\label{corresp}
\begin{array}{c|c}
{\text Linear \, combination \, of \, words} & {\text Line\, defect\, in \,4d}\cr
1 & \hat L_0\cr
Tr\Bigl(M_1^2M_2\Bigr)-3 & \hat L_1\cr
 Tr\Bigl(M_2^2M_1\Bigr)-3&\hat L_2\cr
\frac{1}{3}(I_1-I_2)&\hat L_3\cr
\frac{1}{3}(I_1+I_2)& \hat L_4\cr
\frac{1}{3}(J_1+J_2)& \hat L_5\cr
\frac{1}{3}(J_1-J_2)& \hat L_6\cr
\frac{1}{3}(\tilde J_1+\tilde J_2)& \hat L_9\cr
\frac{1}{3}(\tilde J_1-\tilde J_2)& \hat L_{10}\cr
Tr(M_1^2M_2M_1^2M_2)& \hat L_7\cr
Tr(M_2^2M_1M_2^2M_1)& \hat L_8\cr
\end{array}
\ee
where we denote
$$I_1=Tr\Bigl(M_2^2 M_1 M_2^{-1} M_1^{-1}\Bigr),\quad
I_2=Tr\Bigl(M_1^2 M_2 M_1^{-1} M_2^{-1}\Bigr)$$
$$J_1:=Tr\Bigl(M_2 M_1^2 M_2^2 M_1 M_2^{-1} M_1^{-1}\Bigr),\quad
J_2:=Tr\Bigl(M_1 M_2^2 M_1^2 M_2 M_1^{-1} M_2^{-1}\Bigr)$$
$$\tilde J_1:=Tr\Bigl(M_1^2 M_2 M_1^2 M_2 M_1^{-1} M_2^{-1}\Bigr),\quad
\tilde J_2:=Tr\Bigl(M_2^2 M_1 M_2^2 M_1 M_2^{-1} M_1^{-1}\Bigr)$$
In terms of generators of traces:
$$Tr(M_1^2M_2M_1^2M_2)=R_1^2+6R_1-2R_2+3,\quad
Tr(M_2^2M_1M_2^2M_1)=R_2^2+6R_2-2R_1+3$$
$$\frac{1}{3}(I_1+I_2)=R_1R_2-4(R_1+R_2)+2,\quad \frac{1}{3}(I_1-I_2)=L$$
$$ \frac{1}{3}(J_1+J_2)=R_1R_2(R_1+R_2)-6(R_1^2+R_2^2)-10R_1R_2+2$$
$$\frac{1}{3}(J_1-J_2)=(R_1+R_2)L$$
$$\frac{1}{3}(\tilde J_1+\tilde J_2)=\half R_1R_2(R_1+R_2)+\half (R_2-R_1)L-
2(R_1^2+R_2^2)-R_1R_2-7(R_1+R_2)+2$$
$$\frac{1}{3}(\tilde J_1-\tilde J_2)=-\half R_1R_2(R_2-R_1)-\half (R_1+R_2)L+
2(R_2^2-R_1^2)-3L+(R_1-R_2)$$

Let us define
$$\mathcal{L}_0=\hat L_0,\quad \mathcal{L}_1=\hat L_1+\hat L_2,\quad \mathcal{L}_2=\hat L_2-\hat L_1,\quad
\mathcal{L}_3=\hat L_3$$
We can build a nice basis of line defects $\mathcal{L}_j$ recursively by computing OPEs. For example:
$$\mathcal{L}_1\cdot \mathcal{L}_1=\mathcal{L}_4,\quad 
\mathcal{L}_2\cdot \mathcal{L}_2=\mathcal{L}_5,\quad
\mathcal{L}_1\cdot \mathcal{L}_2=\mathcal{L}_6$$
$$\mathcal{L}_2\cdot \mathcal{L}_3=\mathcal{L}_7,\quad\mathcal{L}_1\cdot \mathcal{L}_3=\mathcal{L}_8$$
where
$$\mathcal{L}_4=\hat L_7+\hat L_8+2\hat L_4+4(\hat L_1+\hat L_2)-10\hat L_0,\quad
\mathcal{L}_5=\hat L_7+\hat L_8-2\hat L_4-4(\hat L_1+\hat L_2)+10\hat L_0$$
$$\mathcal{L}_6=\hat L_8-\hat L_7-8(\hat L_2-\hat L_1),\quad
\mathcal{L}_7=2\hat L_9-\hat L_5-2(\hat L_7+\hat L_8)-8\hat L_4+4(\hat L_1+\hat L_2)+10\hat L_0,\quad \mathcal{L}_8=\hat L_6$$

% Below we give some examples, where by `level' we mean the number of letters
 %in a given word.
 
 %At level 5:
 %$$Tr\Bigl(M^2_{P_0}(v_2)M_{P_0}(v_1)M^{-1}_{P_0}(v_2)M^{-1}_{P_0}(v_1)\Bigr)=
 %{3\over 2}R_1R_2+{3\over 2}L-6(R_1+R_2)+3$$
 %$$Tr\Bigl(M^2_{P_0}(v_1)M_{P_0}(v_2)M^{-1}_{P_0}(v_1)M^{-1}_{P_0}(v_2)\Bigr)=
 %{3\over 2}R_1R_2-{3\over 2}L-6(R_1+R_2)+3$$
 
%At level 8:
%$$J_1:=Tr\Bigl(M_{P_0}(v_2)M^2_{P_0}(v_1)M^2_{P_0}(v_2)M_{P_0}(v_1)M^{-1}_{P_0}(v_2)M^{-1}_{P_0}(v_1)\Bigr)$$
%$$J_2:=Tr\Bigl(M_{P_0}(v_1)M^2_{P_0}(v_2)M^2_{P_0}(v_1)M_{P_0}(v_2)M^{-1}_{P_0}(v_1)M^{-1}_{P_0}(v_2)\Bigr)$$

%$$ J_1+J_2=3 R_1R_2(R_1+R_2)-18(R_1^2+R_2^2)-30R_1R_2+6$$
%$$J_1-J_2=3(R_1+R_2)L$$

In the Yin-Yin chart, i.e.when we choose the Yin-type spectral network in each
of the two triangles, we find

\be \label{traces-gen}  R_1=\left(r_2+{1\over r_2}+2\right){(1+s)^3\over s}, \,
R_2=\left(r_2+{1\over r_2}+2\right){(1+s)^3\over s^2}, \, 
L=\left({1\over r_2}-r_2\right)\left(r_2+{1\over r_2}+2\right){(1+s)^6\over s^3}\ee
Note that there are two basic properties of the vevs of line defects corresponding to the traces. Substitution
$M_1\leftrightarrow M_2$ in a word amounts\footnote{This follows from the definition
of $r_2$ and $s$ as abelian holonomies around certain cycles - exchange of the
punctures $v_1 \leftrightarrow v_2$ reverses the orientation of these cycles.}  to $r_2\leftrightarrow r_2^{-1}$ and
$s \leftrightarrow s^{-1}$ in the vev. Substitution
$M_1\leftrightarrow M_1^{-1}, M_2\leftrightarrow M_2^{-1}$ in a word has similar effect
on the vev.

The generators transform as
$$R_1 \leftrightarrow R_2,\quad L\mapsto -L\quad \text{under}\quad r_2 \leftrightarrow r_2^{-1},\, s\leftrightarrow s^{-1}.$$

\begin{figure}
\includegraphics[width=\textwidth]{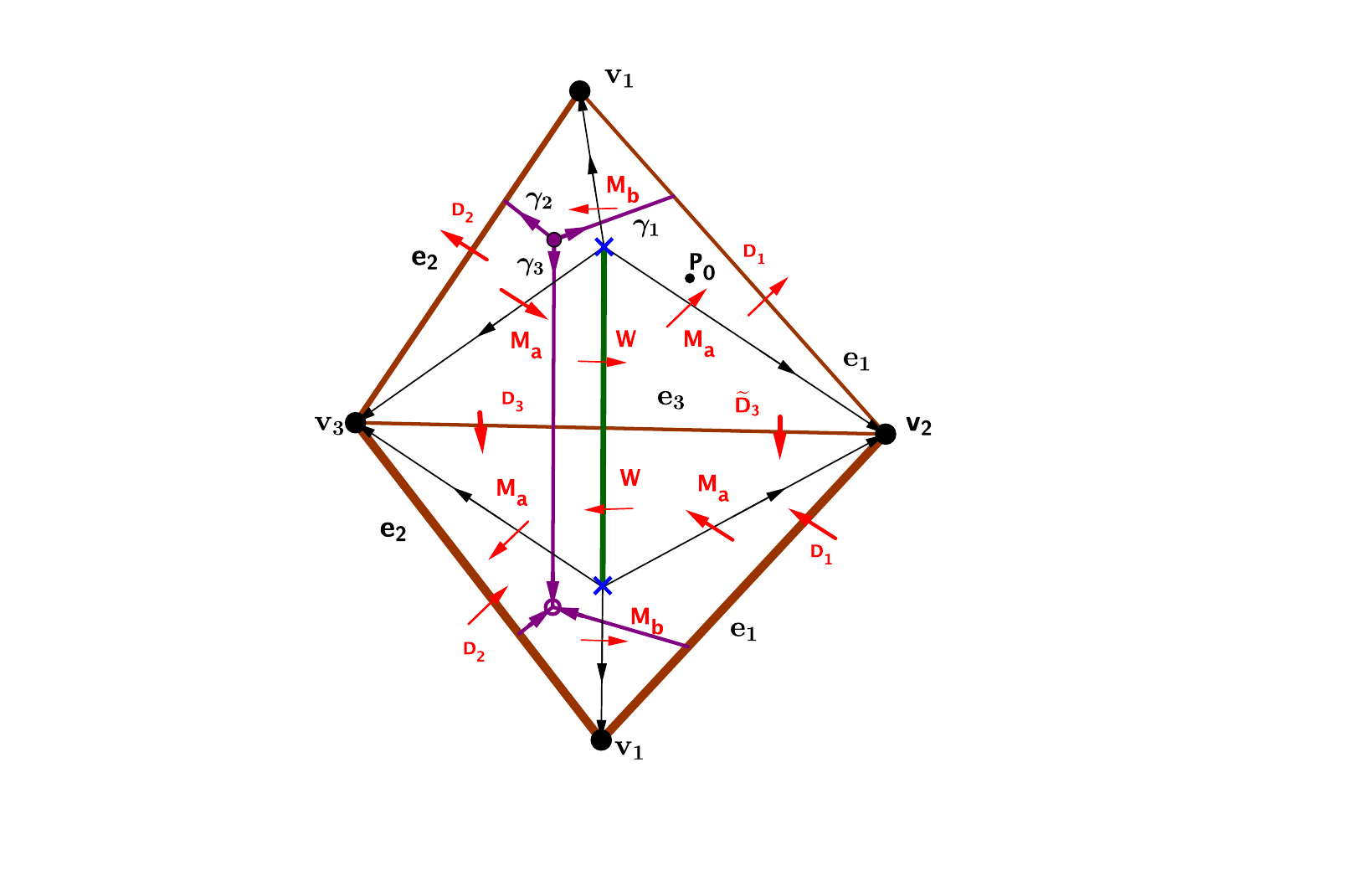}
\vspace{-2cm}
\caption{Web $W_1$}
\label{fig:web1}
\end{figure}

\begin{figure}
%\centering
\includegraphics[width=\textwidth]{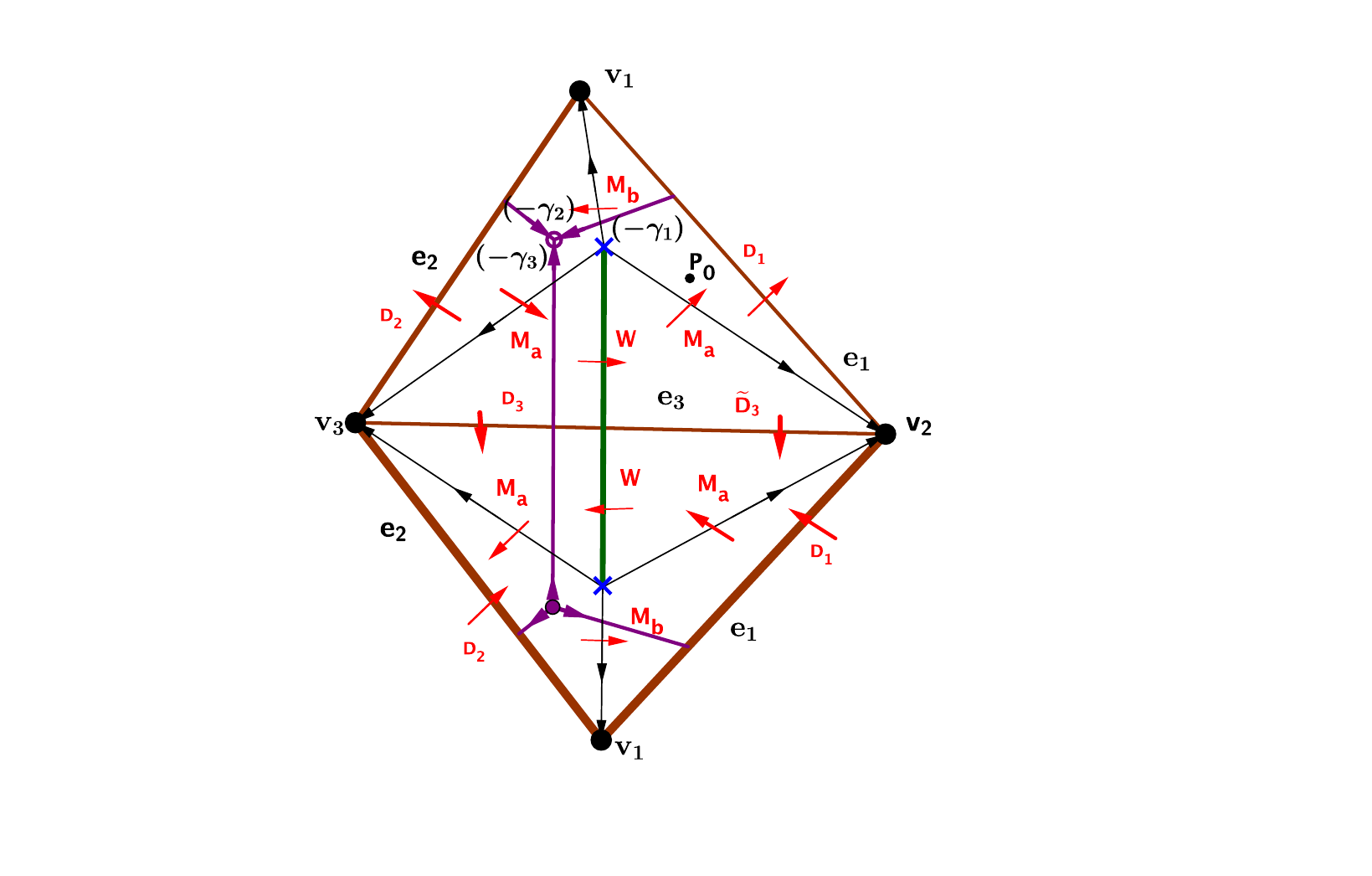}
\vspace{-2cm}
\caption{Web $W_2$}
\label{fig:web2}
\end{figure}

%\
%\begin{figure}
%\centering
%\includegraphics[width=\textwidth]{web5.pdf}
%\vspace{-2cm}
%\caption{Web 6}
%\label{fig:web6}
%\end{figure} \begin{figure}
%\centering

\begin{figure}
%\centering
\includegraphics[width=\textwidth]{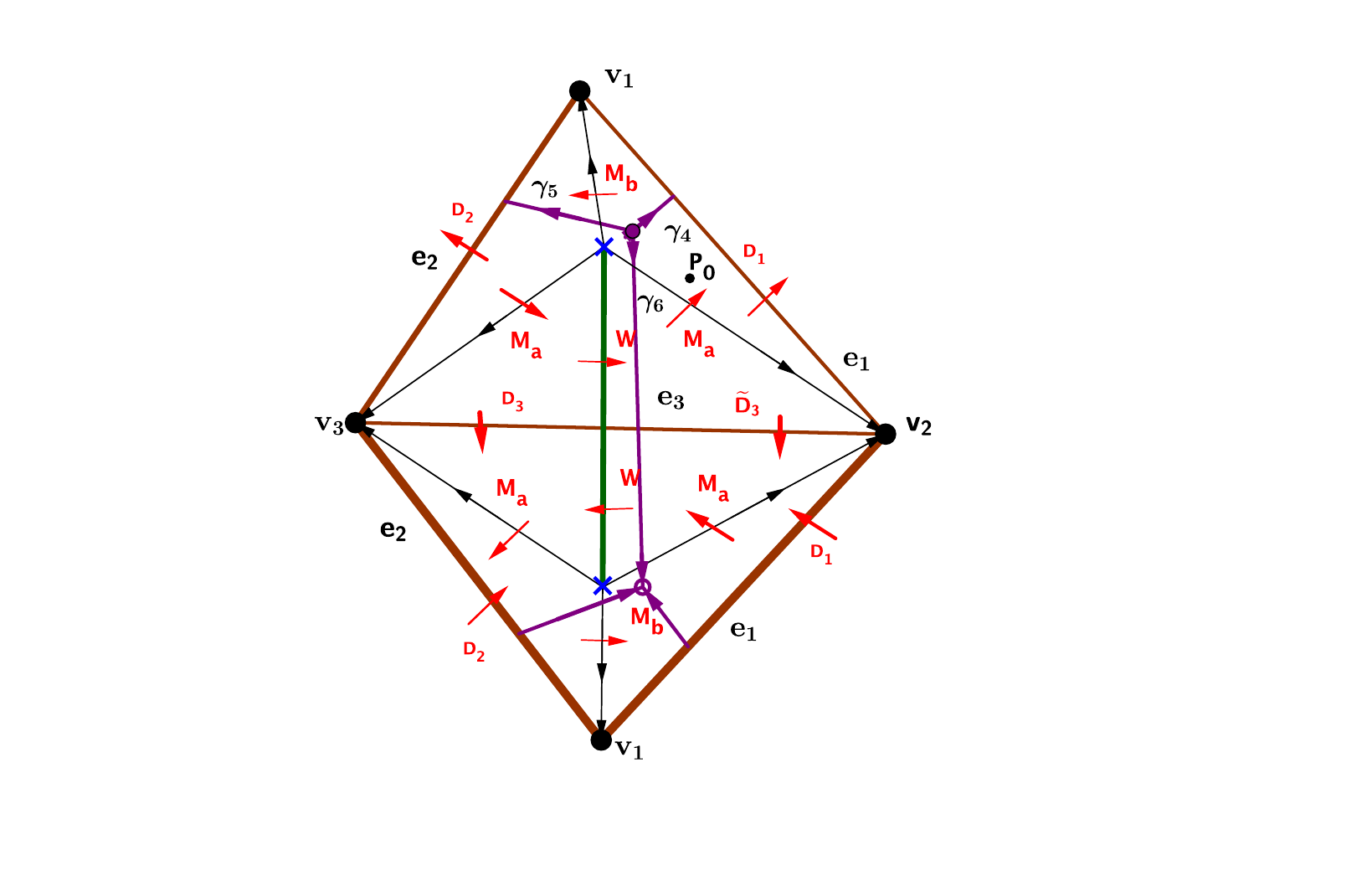}
\vspace{-2cm}
\caption{Web $W_3$}
\label{fig:web3}
\end{figure}

As argued in \cite{Xie,Xie2}, the space of line defects in 4d theory also includes defects
corresponding to webs. The two examples of webs are shown in
Figure \ref{fig:web1} and Figure \ref{fig:web2}. They give line defects with the following vevs:
\be \label{webs-gen}
W_1=6-R_1,\quad W_2=6-R_2.\ee
We computed these from the following gauge-invariant expressions:
$$W_1={web_1\over det M_{\gamma_1}}\quad web_1=\epsilon_{i_1 j_1 k_1}\epsilon^{i_2 j_2 k_2}\left(M_{\gamma_1}\right)^{i_1}_{i_2}\,\left(M_{\gamma_2}\right)^{j_1}_{j_2}\,\left(M_{\gamma_3}\right)^{k_1}_{k_2}$$

$$W_2=web_2 \, det M_{\gamma_1}\quad
web_2=\epsilon_{i_1 j_1 k_1}\epsilon^{i_2 j_2 k_2}\left(M^{-1}_{\gamma_1}\right)^{i_1}_{i_2}\,\left(M^{-1}_{\gamma_2}\right)^{j_1}_{j_2}\,\left(M^{-1}_{\gamma_3}\right)^{k_1}_{k_2}$$
where
$$M_{\gamma_1}=M_b(r_1)^{-1}\, D_{(1)}\, M_b(r_2)^{-1},\quad
M_{\gamma_2}=D_{(2)},\quad M_{\gamma_3}=M_a(r_1)\,D_{(3)}\,M_a(r_2).$$
Note that $W_1$ and $W_2$ are linear combination of traces of holonomies.
It is also interesting that another web $W_3$ (see Figure \ref{fig:web3})
gives the same vev as $W_1$  only at the conformal point. Away from the conformal point the vev of $W_3$ differs as function of masses from the vev of $W_1.$

\begin{figure}
%\centering
\includegraphics[width=\textwidth]{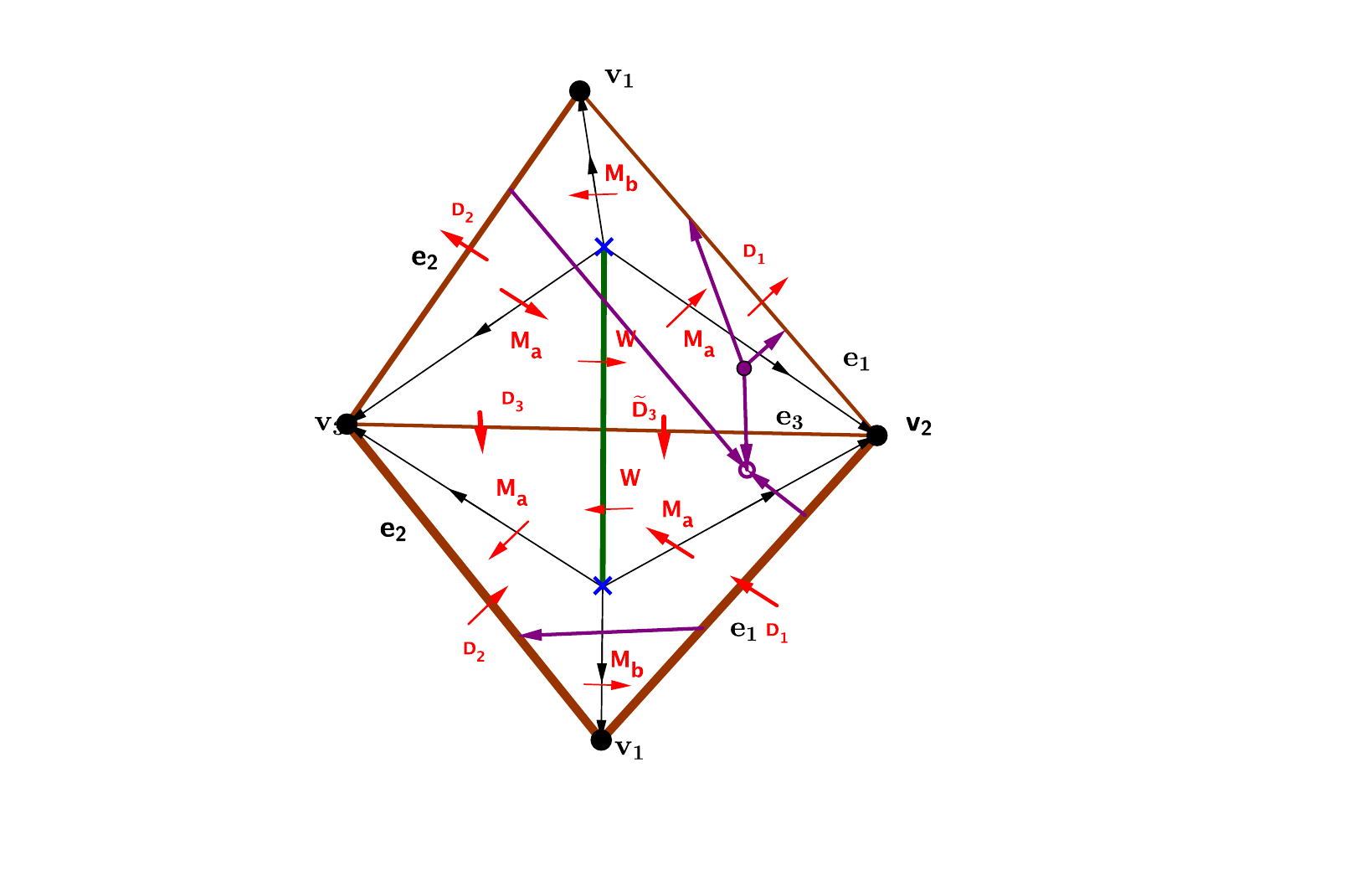}
\vspace{-2cm}
\caption{Web $W_4$}
\label{fig:web4}
\end{figure}

Two more examples of basic webs are 
$W_4$ (see Figure \ref{fig:web4}) and $W_5$ obtained from $W_4$
by reversing the arrows. They give vevs that are again linear combinations of traces of holonomies;
$$W_4=6-R_2,\quad W_5=6-R_1.$$

%It may seem that one should only consider non-contractible webs, like $W_1.\ldots, W_5$ discussed above. However, a contractible web $W_6$ shown in Figure \ref{fig:web6}, as well web $W_7$ obtained from $W_6$ by reversing the arrows, both give non-trivial vevs
%$$W_6=6s, \quad W_7={6\over s}.$$
%In fact $W_6=det D_{(2)}$ where $D_{(2)}$ is a matrix of crossing edge $e_2.$
%The point is that, by specifying spectral network on $C$, 
%we have fixed trivialization of the
%$GL(3)$ bundle $\mathbb{V}$ on $C$ in such way that $lim_{z_1\mapsto z_2}det \left(\mathbb{V}^*_{z_1}\otimes \mathbb{V}_{z_2}\right)=det D_{(j)}$ if $z_1,z_2$ are separated by
%the edge $e_j.$

\begin{figure}
%\centering
\includegraphics[width=\textwidth]{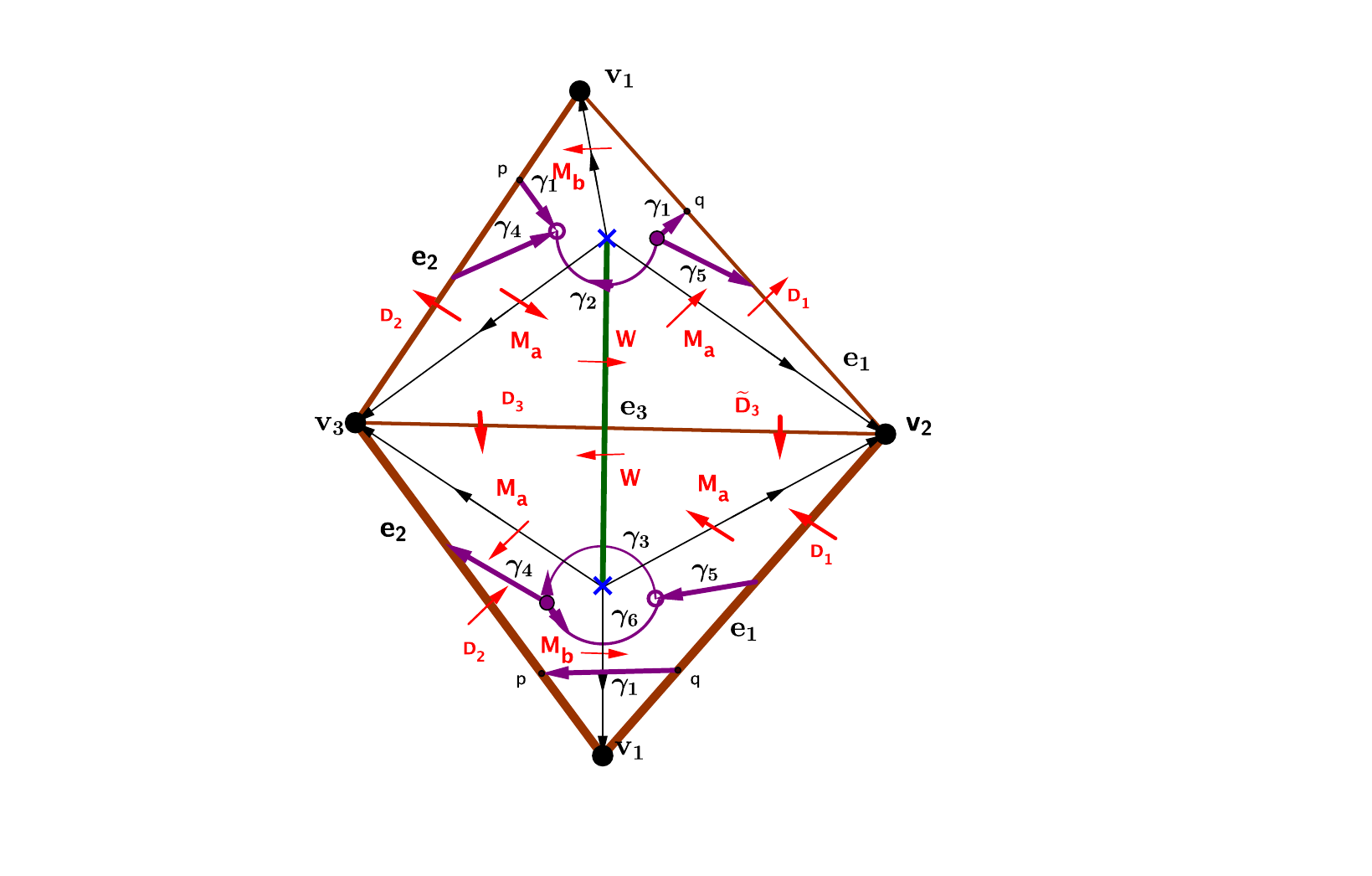}
\vspace{-2cm}
\caption{More complicated web-like structure $\mathbf{W}$}
\label{fig:web-high}
\end{figure}

\begin{figure}
%\centering
\includegraphics[width=\textwidth]{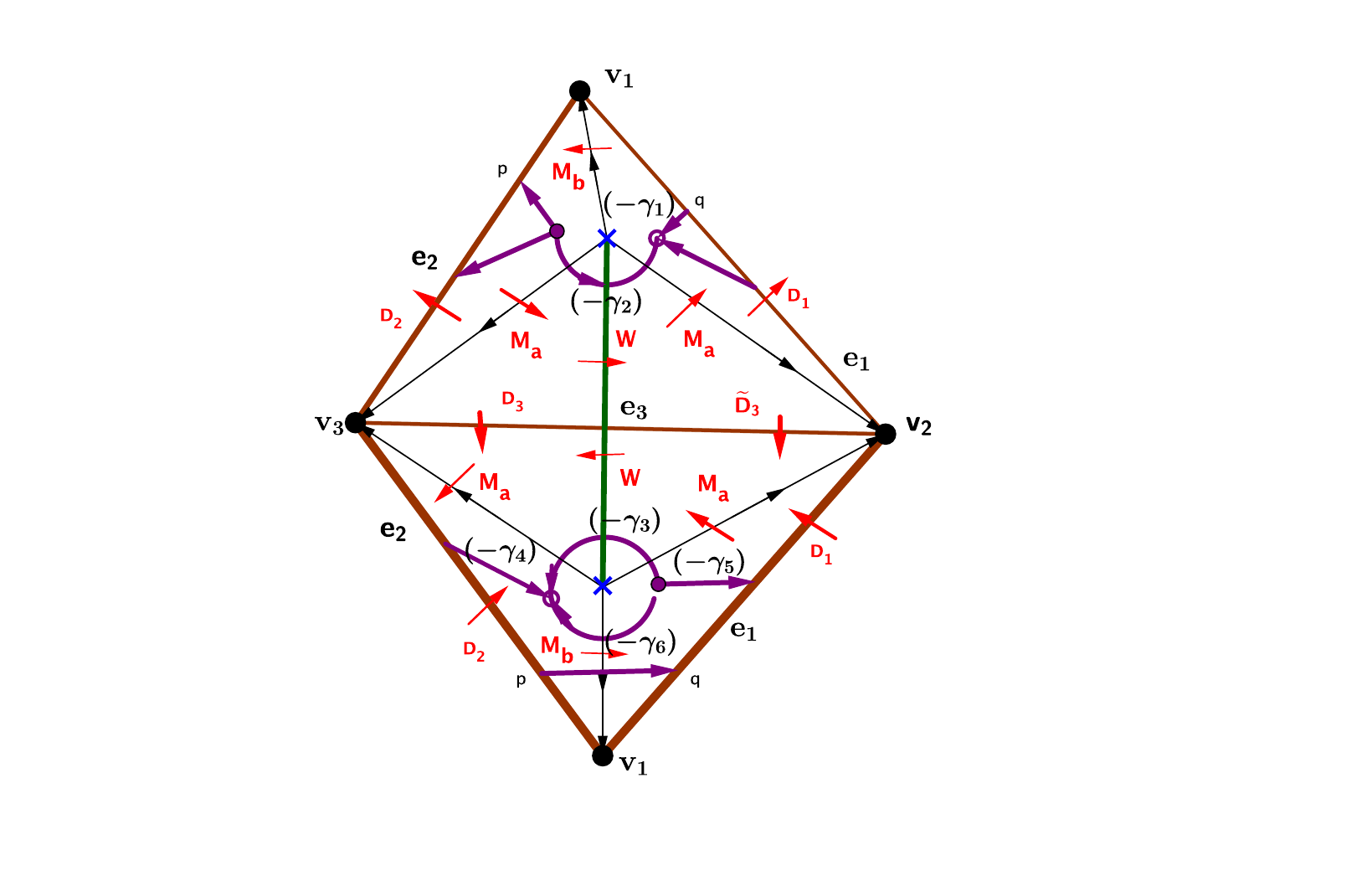}
\vspace{-2cm}
\caption{More complicated web-like structure with $\overline{\mathbf{W}}$}
\label{fig:web-high_ii}
\end{figure}

\begin{figure}
%\centering
\includegraphics[width=\textwidth]{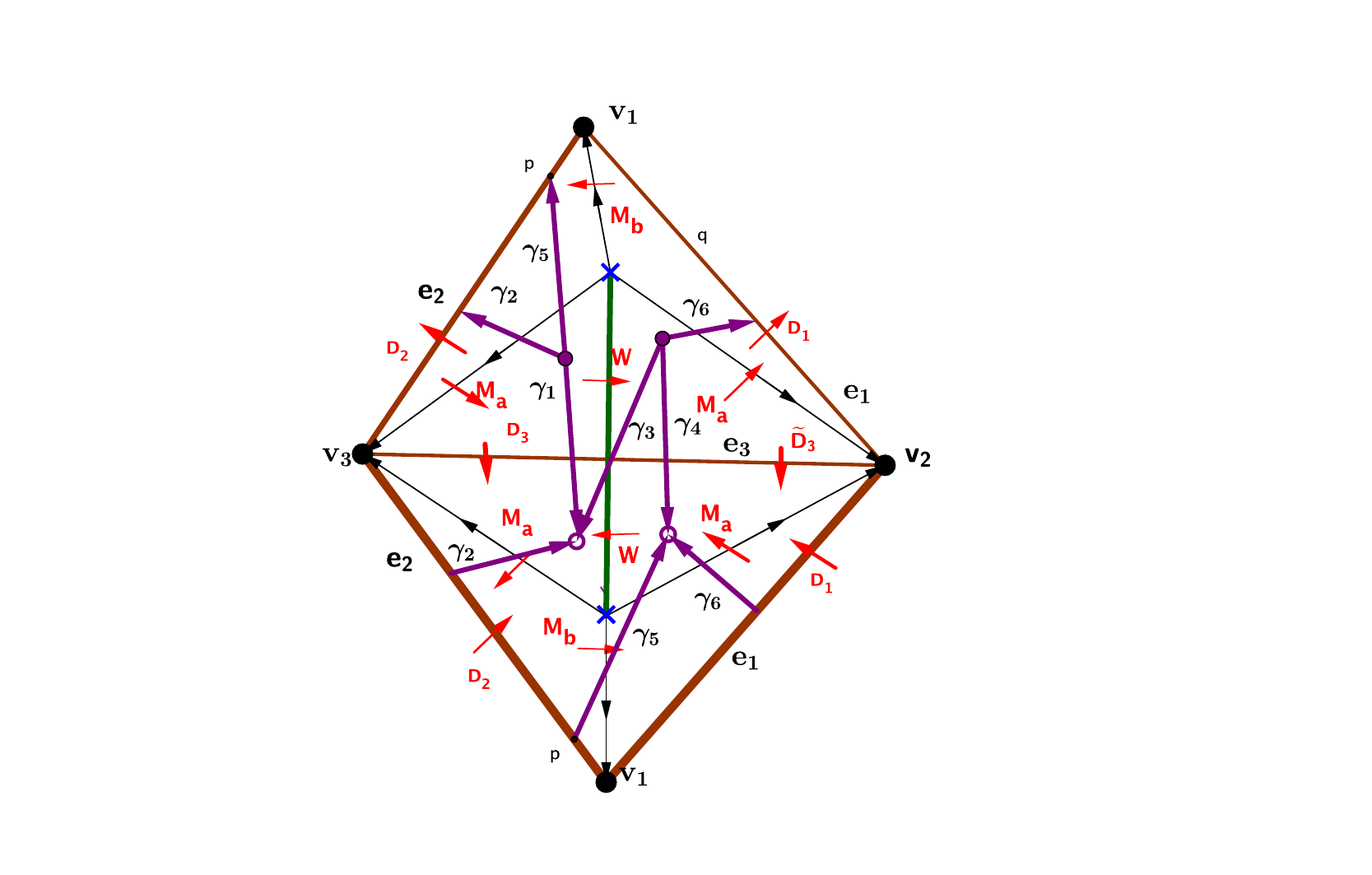}
\vspace{-2cm}
\caption{More complicated web-like structure $\widehat{\mathbf{W}}$}
\label{fig:web-high-new}
\end{figure}

There are line defects arising from more complicated web-like structures. For example, 
the structure depicted in Figure \ref{fig:web-high} gives the vev
$$\mathbf{W}={\eps^{i_1 j_1 n}\, \eps_{i_2 j_2 t}\,\eps^{p_1 p_2 m}\, \eps_{q_1 q_2 s}\,
\left(U^{(\gamma_1)}\right)_{i_1}^{i_2}\, \left(U^{(\gamma_2)}\right)_{j_1}^{j_2}\, 
\left(U^{(\gamma_6)}\right)_{p_1}^{q_1}\,\left(U^{(\gamma_3)}\right)_{p_2}^{q_2}\,
\left(U^{(\gamma_5)}\right)_{n}^{s}\,\left(U^{(\gamma_4)}\right)_{m}^{t}\over det\, U^{(\gamma_4)} \, det \,U^{(\gamma_5)}}$$
where using traffic rule we computed
$$U^{(\gamma_1)}=D_{(2)}^{-1}M^{-1}_b(r_1)D_{(1)},\quad 
U^{(\gamma_2)}=M^{-1}_a(r_2)W^{-1}(r_2)M^{-1}_a(r_2),\quad
U^{(\gamma_3)}=M^{-1}_a(r_1)W^{-1}(r_1)M^{-1}_a(r_1),$$
$$U^{(\gamma_4)}=D_{(2)}^{-1},\quad U^{(\gamma_5)}=D_{(1)},\quad
U^{(\gamma_6)}=M_b(r_1)$$
So that at the conformal point the vev is 
$$\mathbf{W}=10-{8\over r_2}.$$
Reversing all the arrows gives another web, shown in Figure \ref{fig:web-high_ii},
which gives the vev
$$\overline{\mathbf{W}}=10-8 r_2.$$
We see that $\mathbf{W}$ and $\overline{\mathbf{W}}$ are independent as functions
of $(r_2, s)$ from the traces of holonomies.
Let us give another example of more complicated web-like structure (see Figure \ref{fig:web-high-new})
$$\widehat{\mathbf{W}}=-21-\frac{3}{r_2}-r_2+5 r_2^2-\frac{5}{s}-\frac{1}{r_2 s}-\frac{3 r_2}{s}+\frac{r_2^2}{s}-3 s-\frac{3 s}{r_2}+7 r_2 s+7 r_2^2 s+s^2-\frac{s^2}{r_2}+5 r_2 s^2+3 r_2^2 s^2$$
that is also independent of the traces of holonomies. 

Note that we could not find a web-like structure, nor basic neither more complicated, which gives just linear dependence in $s$ without any dependence
on $r_2$. Hence we conclude that we do not get all Functions of $(r_2,s)$ from traces of holonomies and web-like structures.

We finish this section with a comment about the Yang-Yang chart. Requiring that all eigenvalues of monodromies around punctures
are 1, gives
$$x_1=x_2=x_3,\quad y_1=y_2=y_3,\quad z_1=z_2=z_3,\quad r_1r_2=1$$
and relation
\be \label{r2ss2}6-{4\over r_2}-4 r_2+{2\over s}-{2\over r_2 s}+2 s-{2 s\over r_2}+{1\over s_2}+
{2 s\over s_2}-{2 r_2 s\over s_2}+s_2+{2 s_2\over s}-{2 r_2 s_2\over s}=0\ee
where\footnote{In the Yin-Yin chart $s_2=1$.}
$$s_2={z_2\over x_2},\quad s_1={y_2\over x_2}.$$
So that in the Yang-Yang chart all traces (at the conformal point) are functions of $s_1,s_2$ with $r_2$ determined from
the relation (\ref{r2ss2}). 

For example, the generators of the traces are expressed in this chart as
\be\label{f1-s1s2}
R_1=\frac{2(s_1+1) (s_1+s_2)(s_2+1+2s_1)}{s_1 s_2}
\ee
\be \label{f2-s1s2}
R_2=\frac{2(s_1+1) (s_1+s_2)(s_1s_2+2s_2+s_1)}{s_1^2 s_2}
\ee
$$L=\frac{4 (s_1+1)^2 (s_1+s_2)^2 \left(s_1^2 (-4 r_2+2 s_2+2)+s_1 \left((6-8 r_2) s_2+s_2^2+1\right)+2 s_2 (-2 r_2 s_2+s_2+1)\right)}{s_1^3 s_2^2}$$

\section{Relation with Fock-Goncharov variables}
\label{sec:fg}
\begin{figure}
\begin{center}
\includegraphics[width=1.1\textwidth]{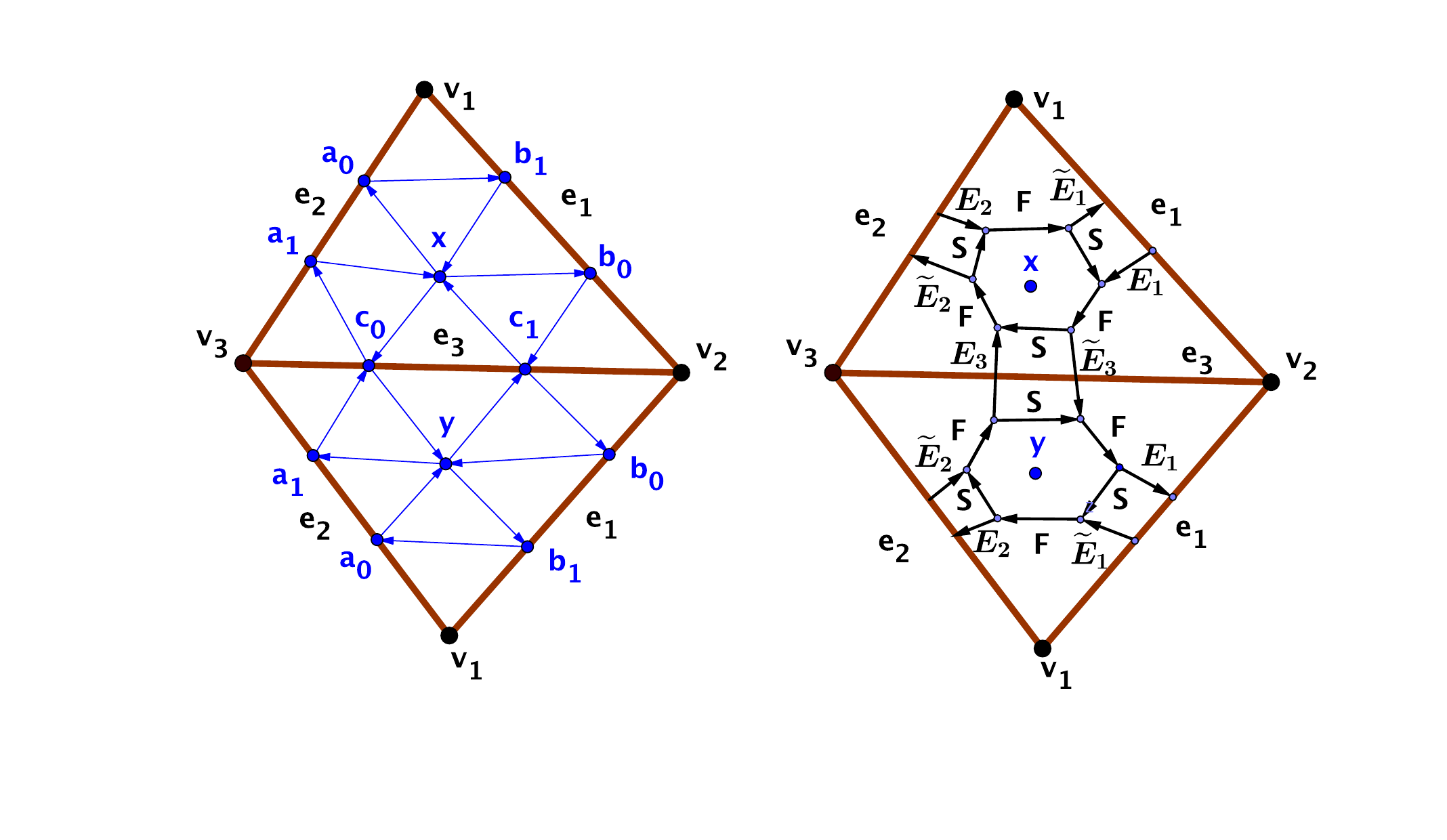}
\caption{Fock-Goncharov variables and monodromy graph for $A_2$ theory on $\mathbb{P}^1/\{v_1,v_2,v_3\}$}
\label{fg_var}
\end{center}
\end{figure}
\begin{figure}
\begin{center}
\includegraphics[width=1.2\textwidth]{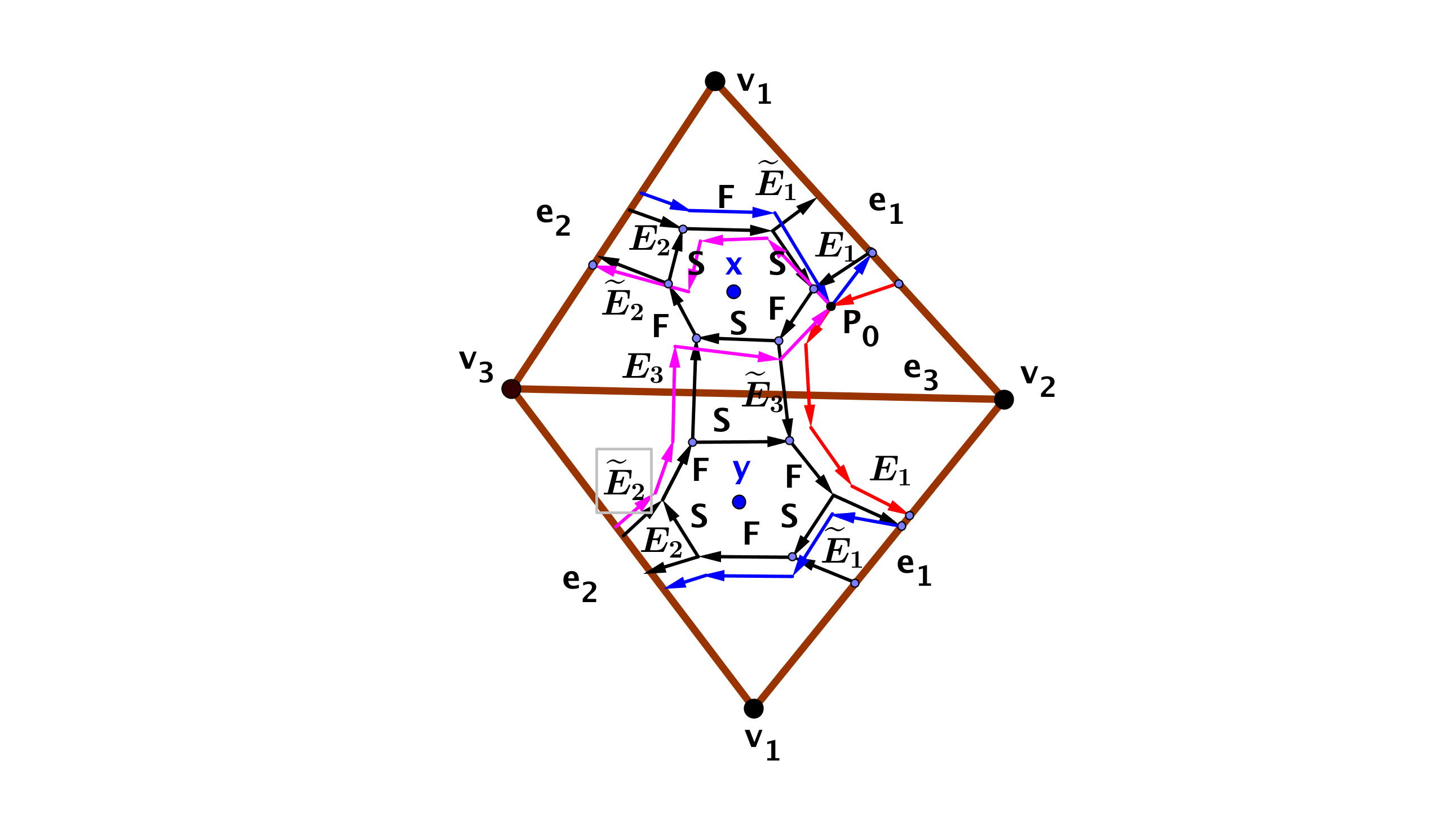}
\vspace{-2cm}
\caption{Monodromies around punctures in Fock-Goncharov variables for $A_2$ theory on $\mathbb{P}^1/\{v_1,v_2,v_3\}$}
\label{fig:monod_fg}
\end{center}
\end{figure}

The Fock-Goncharov variables and monodromy graph \cite{FG} for moduli space of $PGL(3,\mathbb{C})$ connections on $\mathbb{P}^1/\{v_1,v_2,v_3\}$ are depicted in Figure 4. The non-zero Poisson brackets are
\be \label{non-zero-fg}\{x,a_0\}=xa_0,\quad \{x,a_1\}=-xa_1,\quad\{x,b_0\}=xb_0,\quad \{x,b_1\}=-xb_1,\quad
\{x,c_0\}=xc_0,\quad \{x,c_1\}=-xc_1\ee
$$\{y,a_0\}=-ya_0,\quad \{y,a_1\}=ya_1,\quad\{y,b_0\}=-yb_0,\quad \{y,b_1\}=yb_1,\quad
\{y,c_0\}=-yc_0,\quad \{y,c_1\}=yc_1$$
Let us write monodromies around punctures (see Figure \ref{fig:monod_fg}) with the same base point $P_0$
that we used in computing the corresponding monodromies in spectral network set up. 
$$M^{FG}_{P_0}(v_1)=S\,F(x)\, E_2\, F(y)\, S\, E_1^{-1}$$
$$M^{FG}_{P_0}(v_2)=E_1 \, F(y)\, \tilde{E}_3\, F(x)$$
$$M^{FG}_{P_0}(v_3)=F^{-1}(x)\, S\,  E_3\, F(y)\,\tilde{E}_2\, S\, F^{-1}(x)\,S$$
where
$$S=\begin{pmatrix}0&0&-1\cr0&1&0\cr-1&0&0\cr\end{pmatrix},\quad 
F(x)=\begin{pmatrix} x&0&0\cr x&x&0\cr x&x + 1&1\end{pmatrix},\quad
E_1=\begin{pmatrix}b_0b_1 &0 &0\cr 0& b_1 &0\cr 0&0&1\end{pmatrix},\quad
\tilde{E}_1=\begin{pmatrix}b_0b_1 &0 &0\cr 0& b_0 &0\cr 0&0&1\end{pmatrix},\quad
$$
$$E_2=\begin{pmatrix}a_0a_1 &0 &0\cr 0& a_1 &0\cr 0&0&1\end{pmatrix},\quad
\tilde{E}_2=\begin{pmatrix}a_0a_1 &0 &0\cr 0& a_0 &0\cr 0&0&1\end{pmatrix},\quad
E_3=\begin{pmatrix}c_0c_1 &0 &0\cr 0& c_1 &0\cr 0&0&1\end{pmatrix},\quad
\tilde{E}_3=\begin{pmatrix}c_0c_1 &0 &0\cr 0& c_0 &0\cr 0&0&1\end{pmatrix},\quad
$$
%Monodromies satisfy
%$$M_{P_0}^{FG}(v_3)\,M^{FG}_{P_0}(v_1)\, M^{FG}_{P_0}(v_2)=a_0a_1c_0c_1y^2\, \Id$$
To write down eigenvalues (defined up to a scale) of these monodromy matrices,
it is convenient to introduce the following combinations of FG variables
$$\Upsilon^{(1)}_1=a_0b_1, \, \Upsilon^{(1)}_2=a_1b_0,\,\Upsilon^{(2)}_1=b_0c_1, \, \Upsilon^{(2)}_2=b_1c_0,\,\Upsilon^{(3)}_1=a_1c_0, \, \Upsilon^{(3)}_2=a_0c_1,\, f=xy.$$
Note the relation:
\be \label{rel-ups}\Upsilon_1^{(1)}\,\Upsilon_1^{(2)}\,\Upsilon_1^{(3)}\,=\,
\Upsilon_2^{(1)}\,\Upsilon_2^{(2)}\,\Upsilon_2^{(3)}
\ee

We find the eigenvalues 
$$\bigl(n^{(a)}_1, n^{(a)}_2, n^{(a)}_3\bigr):=
Eigen\Bigl(M^{FG}_{P_0}(v_a)\Bigr)
=\Lambda^{(a)}\bigl(1, \xi^{(a)}_1, \xi^{(a)}_2\bigr)\quad a=1,2,3.$$
where $\Lambda^{(a)}$ are arbitrary scale factors and
\be \label{def-xi}\xi^{(a)}_1=f \Upsilon^{(a)}_1, \quad \xi^{(a)}_2=f \Upsilon^{(a)}_1 \Upsilon^{(a)}_2.\ee
As a consequence of (\ref{rel-ups}):
\be \label{f-expr}f^3={\left(\xi^{(1)}_2\,\xi^{(2)}_2\, \xi^{(3)}_2\right)^2\over
\xi^{(1)}_1\,\xi^{(2)}_1\, \xi^{(3)}_1}.\ee

For each puncture let us compute the two invariant (under rescaling and Weyl group action) combinations
\be \label{def-rho}\rho^{(a)}_1={\Bigl(n^{(a)}_1+n^{(a)}_2+n^{(a)}_3\Bigr)^3\over n^{(a)}_1\, n^{(a)}_2\, n^{(a)}_3}={\Bigl(1+\xi^{(a)}_1+\xi^{(a)}_2\Bigr)^3\over \xi^{(a)}_1\,\xi^{(a)}_2}\ee
$$\rho^{(a)}_2={\Bigl(n^{(a)}_1+n^{(a)}_2+n^{(a)}_3\Bigr)\Bigl(n^{(a)}_1n^{(a)}_2+
n^{(a)}_2n^{(a)}_3+n^{(a)}_1n^{(a)}_3\Bigr)\over n^{(a)}_1\, n^{(a)}_2\, n^{(a)}_3}
={\Bigl(1+\xi^{(a)}_1+\xi^{(a)}_2\Bigr)\Bigl( \xi^{(a)}_1+\xi^{(a)}_2+\xi^{(a)}_1\xi^{(a)}_2\Bigr)\over \xi^{(a)}_1\,\xi^{(a)}_2}
$$
Now it is clear how to explicutly relate Fock-Goncharov and spectral network variables.
We first compute $\rho^{(a)}_1, \rho^{(a)}_2$ from spectral network, i.e. as functions
of masses $k_i.$ Then we find $\xi^{(a)}_1,\xi^{(a)}_2$ from $\rho^{(a)}_1,\rho^{(a)}_2,
$ compute $f$ from (\ref{f-expr}) and determine $\Upsilon^{(a)}_1,\Upsilon^{(a)}_2$
from (\ref{def-xi}). Finally, let us recall the relation between FG variables and $f,\Upsilon^{(a)}_{1},\Upsilon^{(a)}_2$ (\ref{rel-ups}) and Poisson brackets (\ref{non-zero-fg}). This allows
to write an explicit map
\be \label{fg-sn-map}a_0=s,\, x=r_2,\, y={f\over r_2},\, b_1={\Upsilon_1^{(1)}\over s},\, c_1={\Upsilon_2^{(3)}\over s},\,c_0={s\,\Upsilon_2^{(2)}\over \Upsilon_1^{(1)}},\, a_1={\Upsilon_1^{(3)}\Upsilon_1^{(1)}\over s \Upsilon_2^{(2)}},\, b_0={s\, \Upsilon_1^{(2)}\over
\Upsilon_2^{(3)}}\ee

The only technical difficulty arises in solving for $\xi^{(a)}_1,\xi^{(a)}_2$ from $\rho^{(a)}_1,\rho^{(a)}_2.$ One may first determine $\mu^{(a)}=\xi^{(a)}_1\xi^{(a)}_2$ and
$\nu^{(a)}=\xi^{(a)}_1+\xi^{(a)}_2+1$ and then solve quadratic equation to find $\xi$'s
from $\mu$ and $\nu.$ Note that from (\ref{def-rho})
$$\mu^{(a)}={\left(\nu^{(a)}\right)^3\over  \rho_1^{(a)}}$$
and $\nu^{(a)}$ is a solution of a cubic equation for each puncture $a=1,2,3$
$$\left(\nu^{(a)}\right)^3-\rho_2^{(a)}\left(\nu^{(a)}\right)^2+\rho_1^{(a)}\nu^{(a)}-\rho_1^{(a)}=0.$$
%{\bf Any solution of this cubic equation is good for us?}
%\vspace{4cm}
\section{Acknowledgments}
I am very grateful to Greg Moore for many 
valuable advices without which this work would not be possible. I 
would like to express my thanks to the Aspen Center for Physics for
hospitality during July 2013.

\section{Appendix}
Let us consider a deviation from the conformal point in the base of $\mathcal{M}$ parametrized as
$$f_1^1=n+2;\quad f_1^2=n^2+2;\quad f_1^3=n^3+2;\quad f_2^j=3;\quad
f_3^1=n^{-1}+2;\quad f_3^2=n^{-2}+2;\quad f_3^3=n^{-3}+2.
$$
Below we provide coefficients of the equation (\ref{eq-fiber}) for this simple deviation.
$$g_1=\frac{n+2}{3 n^2};\quad g_3=\frac{-n^2-4 n+5}{3 n^2+15 n};\quad
 u_1=\frac{3 n^3+13 n^2-11 n-5}{3 n^2+15 n};$$
  $$g_2=\frac{-n^2-10 n-25}{9 n^2};\quad g_4=\frac{-3 n^3+3 n^2 u_1-23 n^2+6 n u_1-47 n-35}{9 n^2};$$
  $$ g_6=\frac{-n^5-11 n^4+3 n^3 u_2-38 n^3+6 n^2 u_2-43 n^2-15 n}{9 n^3};\quad u_2=\frac{n^5+8 n^4+10 n^3-24 n^2+5 n}{3 n^3+15 n^2};$$
 $$ g_5=\frac{-n^4-10 n^3-25 n^2}{9 n^3};\quad
  g_7=\frac{2 n^2}{9}-\frac{119}{9 n^2}+\frac{46 n}{9}-\frac{4}{n}+\frac{107}{9};$$ 
  $$g_9=\frac{34 n^4+157 n^3+108 n^2-175 n-124}{9 n^2};$$
   $$g_{10}=\frac{-n^6-58 n^5-171 n^4+91 n^3+380 n^2-53 n-188}{9 n^2};$$
   $$g_{11}=\frac{-15 n^6-100 n^5-119 n^4+226 n^3+293 n^2-198 n-87}{9 n^2};$$
    $$ g_8=\frac{3 n^5+30 n^4+87 n^3+29 n^2-130 n-19}{9 n^2};u_3=\frac{-3 n^5-17 n^4+46 n^2-21 n-5}{3 n^2+15 n};$$
    $$g_{12}=\frac{(n-1)^2 \left(22 n^5+130 n^4+206 n^3-25 n^2-240 n-94\right)}{9 n^2}$$
    We find that for generic $n$ the fiber of $\mathcal{M}$ is a non-singular complex surface in $\mathbb{C}^3.$

\end{document}